\documentclass[useAMS,usenatbibi,usegraphicx]{mn2e}
\usepackage{psfig}

\newcommand{\be}{\begin{equation}}
\newcommand{\ee}{\end{equation}}
\newcommand{\ba}{\begin{eqnarray}}
\newcommand{\ea}{\end{eqnarray}}
\newcommand{\Mc}{{\cal M}}
\newcommand{\Ms}{M_{\odot}}
\newcommand{\m}{\langle}
\newcommand{\M}{\rangle}

\newcommand{\bml}{\begin{mathletters}}
\newcommand{\eml}{\end{mathletters}}

\def\ltsima{$\; \buildrel < \over \sim \;$}
\def\simlt{\lower.5ex\hbox{\ltsima}}
\def\gtsima{$\; \buildrel > \over \sim \;$}
\def\simgt{\lower.5ex\hbox{\gtsima}}
\def\gsim{ \lower .75ex \hbox{$\sim$} \llap{\raise .27ex \hbox{$>$}} }
\def\lsim{ \lower .75ex\hbox{$\sim$} \llap{\raise .27ex \hbox{$<$}} }
\def\msun{\,{\rm M_\odot}}

\title[Gravitational waves from black hole binaries]
{The stochastic gravitational-wave background from massive black hole binary systems: implications for observations with 
Pulsar Timing Arrays}

\author[A. Sesana et al.]{A. Sesana$^{1,2}$, A. Vecchio$^{1,3}$
and C. N. Colacino$^{1}$\\
$^{1}$School of Physics and Astronomy, The University of Birmingham, 
Edgbaston, Birmingham, B15 2TT, UK\\
$^{2}$Center for Gravitational Wave Physics, Penn State University, University Park, PA 16802, USA\\
$^{3}$ Department of Physics and
  Astronomy, Northwestern University, 2145 Sheridan Road, Evanston, IL 60208, USA}

\begin{document}

\date{Received ---}

\maketitle

\begin{abstract}
Massive black hole binary systems, with masses in the range $\sim 10^4-10^{10}\msun$, are among the primary sources of gravitational waves in the frequency window $\sim 10^{-9}\,\mathrm{Hz} - 0.1 \,\mathrm{Hz}$. Pulsar Timing Arrays (PTAs) and the Laser Interferometer Space Antenna ({\it LISA}) are the observational means by which we will be able to observe gravitational radiation from these systems. We carry out a systematic study of the generation of the stochastic gravitational-wave background from the cosmic population of massive black hole binaries. We consider a wide variety of assembly scenarios and we estimate the range of signal strength in the frequency band accessible to PTAs. We show that regardless of the specific model of massive black hole binaries formation and evolution, the characteristic amplitude $h_c$ of the gravitational wave stochastic background at $10^{-8}$ Hz varies by less than a factor of 2. However, taking into account the uncertainties surrounding the actual key model parameters, the amplitude lies in the interval $h_c(f = 10^{-8}\,\mathrm{Hz}) \approx 5\times 10^{-16} - 8\times 10^{-15}$. The most optimistic predictions place the signal level at a factor of $\approx 3$ below the current sensitivity of Pulsar Timing Arrays, but within the detection range of the complete Parkes PTA for a wide variety of models, and of the future Square-Kilometer-Array PTA for all the models considered here. We also show that at frequencies $\simgt 10^{-8}$ Hz the frequency dependency of the generated background follows a power-law significantly steeper than $h_c\propto f^{-2/3}$, that has been considered so far; the value of the spectral index depends on the actual assembly scenario and provides therefore an additional opportunity to extract astrophysical information about the cosmic population of massive black holes. Finally we show that {\it LISA} observations of individual resolvable massive black hole binaries are complementary and orthogonal to PTA observations of a stochastic background from the whole population in the Universe. In fact, the detection of gravitational radiation in both frequency windows will enable us to fully characterise the cosmic history of massive black holes.

\end{abstract}

\section{Introduction}
\label{s:intro}

Massive black hole (MBH) binary systems with masses in the range range $\sim 10^4-10^{10}\msun$ are amongst the primary candidate 
sources of gravitational waves (GWs) at $\sim$ nHz - mHz frequencies (see, e.g., 
Haehnelt 1994; Jaffe \& Backer 2003; Wyithe \& Loeb 2003, Sesana et al. 2004, 
Sesana et al. 2005). MBHs are ubiquitous today in the 
nuclei of nearby galaxies (see, e.g., Magorrian et al. 1998). If MBHs were 
also common in the past (as implied by the notion that many distant galaxies 
harbour active nuclei for a short period of their life, e.g. Haehnelt \& Rees 1993), and if their host 
galaxies experienced multiple mergers during their lifetime, as dictated by 
popular cold dark matter hierarchical cosmologies (e.g. White \& Rees 1978, Peebles 1982), 
then massive black hole binaries (MBHBs) would inevitably form during cosmic history.
Many different scenarios for MBH formation and evolution have been proposed 
(e.g. Volonteri, Haardt \& Madau 2003, Koushiappas, Bullock \& Dekel 2004, 
Begelman, Volonteri \& Rees 2006, Lodato \& Natarajan 2006): the predicted number density of MBHBs and their distribution 
as a function of mass and redshift vary considerably, especially at $z\simgt2$, where current observational constraints
are not tight, but all of them predict the existence of  
MBHBs whose GW emission could be 
observed with present-day and future experiments. 

GWs emitted by MBHBs during their whole coalescence span a frequency range that extends from the 
nHz to the mHz region. The frequency band $\sim 10^{-5}\,\mathrm{Hz} - 1 \,\mathrm{Hz}$ will be probed by the 
{\it Laser Interferometer Space Antenna} ({\it LISA}, Bender et al. 1998); {\it LISA} will observe the last months-to-years of the 
inspiral of two MBHs followed by the catastrophic merger and ringdown phases. In this frequency region, MBHBs are very
bright (in the GW band) resolved sources. The frequency 
window $10^{-9}\,\mathrm{Hz} - 10^{-6} \,\mathrm{Hz}$ is 
accessible by {\em ongoing} (e.g. the Parkes radio-telescope, Manchester 2008) and planned (e.g. the 
Square Kilometer Array (SKA), {\it www.skatelescope.org} 
) Pulsar Timing Array (PTA) experiments. The observable window is set at low frequencies by 
the span of the observations, typically lasting several years, and 
at high frequencies by the time interval between 
successive observations of the same pulsar, typically a week. At these frequencies the most likely signal to be detected 
is a stochastic GW background generated by the incoherent superposition of radiation from the whole cosmic population of MBHBs\footnote{ Stochastic signals generated by the incoherent superposition of deterministic signals from astrophysical sources are usually referred to as {\em foregrounds} (especially in the {\it LISA} community) to distinguish them from the background of gravitational waves produced in the very-early Universe. Here we stick to the naming convention of the radio-pulsar community and we will refer to the stochastic signal from a cosmic population of massive black hole binaries as {\em background}.},
although individual sources could also be observed (see, e.g., Jenet et al., 2004, 2005a). Pulsar timing provides therefore a unique tool to study GWs at very low frequencies. 

GWs affect the propagation of radio signals from the pulsar to the receiver on Earth (e.g. Sazhin 1978, Detweiler 1979, Bertotti et al. 1983), producing a characteristic signature in the time of arrival (TOA) of radio pulses.  In general, one computes the difference between the expected arrival time, according to a given model, and the actual arrival time (e.g. Helling \& Downs 1983, Jenet et al. 2005b); the residuals from this fit carry the physical information about the unmodelled effects in the pulse propagation, including possible GWs. Ongoing and future PTAs are reaching sensitivity 
levels that can lead to the detection of a stochastic background generated by MBHBs. It is therefore essential to 
understand the uncertainties affecting the predictions from different models and to rigorously characterise the statistical
properties of the signal, which in turn affect the data analysis approach.

So far, the stochastic GW background from MBHBs in the frequency window 
$\sim 10^{-9}\,\mathrm{Hz} - 10^{-7}\,\mathrm{Hz}$ has been treated as an isotropic, Gaussian and stationary signal described by a characteristic amplitude $h_c$ that follows a power law $h_c\propto f^{-2/3}$ (e.g. Phinney 2001). Consequently, the relevant data analysis technique to search for such a signal consists in looking for statistically significant correlations in the TOA residuals of different pulsars. Jenet et al. (2006) placed the currently tightest constraint on the level of the GW background, and  showed (Jenet et al. 2005b) that observations of 20 millisecond pulsars spanning 5 years with a timing precision of $\sim 100$ ns would allow the detection of a GW background at the level predicted by several models of MBH assembly (Wyithe \& Loeb 2003, Sesana et al. 2004, Enoki et al. 2004). 

In this paper we carry out a systematic study of the generation of a GW stochastic background from MBHBs in the 
frequency range accessible to PTAs by considering a wide variety of models of MBH assembly. They qualitatively 
encompass the whole range of scenarios currently proposed for MBH formation and quantitatively cover a 
very broad spectrum of predictions. The main results of our analysis can be summarised as follows (the reader who wants to skip all the details and just see the essential results is advised to refer to equation~\ref{hfit} -- with parameter values provided in Table~\ref{tab2} and equations~\ref{fitparameters1}-\ref{fitparameters3} --  and Figs ~\ref{figdrop} and~\ref{detection}). 

\begin{enumerate} 
\item We show that contrary to what has been assumed so far, the generated stochastic background does {\em not} follow a power law $h_c\propto f^{-2/3}$ across the whole
frequency range $\sim 10^{-9}\,\mathrm{Hz} - 10^{-7}\,\mathrm{Hz}$, but the frequency dependence becomes steeper 
above $\approx 10^{-8}$ Hz -- see equation (\ref{hfit}), Table \ref{tab2} and Fig. \ref{figdrop} -- due to the intrinsic {\it discrete nature} of the sources. 
\item The frequency
at which there is a significant deviation from the behaviour $h_c\propto f^{-2/3}$ and the new spectral index both depend on
the actual MBHBs cosmic history, see Table \ref{tab2}. 
\item We estimate the strength of the stochastic GW background 
as a function of the adopted astrophysical model and the uncertainties surrounding the model's parameters, 
see equations~(\ref{fitparameters1})-(\ref{fitparameters3}) and Fig. \ref{figunc}.
Regardless of the specific model of MBHB formation and evolution, the characteristic amplitude $h_c$ of the gravitational wave stochastic background at $10^{-8}$ Hz varies by less than a factor of 2; however, taking into account the uncertainties surrounding the actual key model parameters, the amplitude varies by a factor of $\approx 20$.
\item The most optimistic predictions place the signal level at a factor of $\approx 3$ below the current sensitivity of PTAs, but future surveys are likely to detect the signal predicted by the whole range of astrophysical models considered
here; they can potentially measure the spectral shape of the background in the frequency range $3\times 10^{-9}-3\times 10^{-8}$ Hz 
and therefore provide new information on the MBH population at low redshift, see Fig.~\ref{detection}.
\item PTA observations of a stochastic 
background are complementary and ``orthogonal'' to {\it LISA} observations of 
individual events; in fact, models of MBH assembly that generate large (by over an order of magnitude) 
fluctuations in the number of {\it LISA} events, can provide very similar levels of stochastic backgrounds in the PTA frequency 
range and vice-versa; this is due to the fact that PTAs are especially sensitive to the low-redshift population of MBHBs, 
whereas {\it LISA} will be able to observe also very high-redshift sources, see Fig.~\ref{figskatch}.
\item As a byproduct of our analysis, we have compared 
the observed massive (luminous) galaxy merger rates at $z<1$ derived by a number of authors 
with the merger rates inferred using the extended Press \& Schechter  (EPS) approach adopted in this paper,
and find good agreement.

\end{enumerate}

The paper is organised as follows. In Section 2 we summarise the theory
of the generation of GWs from a population of MBHBs. In Section 3 we introduce four different MBH assembly scenarios that are representative of the formation and evolution of this class of sources and evaluate the total GW signal using a 
semi-analytical approach and Monte-Carlo methods. In Section 4 we derive a more
accurate estimate of the GW stochastic background and introduce a simple mathematical 
model to account for the discrepancy between semi-analytical predictions and
Monte-Carlo simulations. Section 5
is devoted to quantifying the uncertainties in the strength of the background and
in Section 6 we compare our results with those from previous works. Observational prospects are discussed in Section 7, and the main results and pointers to future work are summarised in Section 8. 

\section{gravitational wave backgrounds: key concepts}
In this section we review the key concepts and formulae regarding a GW stochastic background signal produced by the superposition of radiation from a large number of individual sources. For more details, we refer the reader to Phinney (2001), and references therein. Throughout the paper, unless differently specified, we adopt geometrical units in which $c = G = 1$. 

A stochastic background is described in terms of the present-day GW energy density $\rho_{\rm gw}(f)$ per 
logarithmic frequency interval normalized to the critical energy density $\rho_\mathrm{c}$ according to:
\begin{equation}
\Omega_\mathrm{gw}(f) = \frac{1}{\rho_\mathrm{c}}\,\frac{d\rho_{\rm gw}(f)}{d\ln f}\,.
\label{e:Omega}
\end{equation}
Here $f$ is the {\em observed} frequency at the instrument. Let us now consider a population of GW events characterized by a comoving number density per unit redshift $dn/dz$; the events generate energy per logarithmic frequency interval $dE_{\rm gw}/d\ln{f_r}$, where the energy is measured in the source rest-frame at redshift $z$ and $f_r = (1 + z) f$ is the \emph{rest-frame} frequency. The energy spectrum $\Omega_\mathrm{gw}(f)$ of the background produced by the superposition of the radiation from these cosmic events is therefore given by (Phinney, 2001) 
%
%
\begin{equation}
\frac{d\rho_{\rm gw}(f)}{d\ln f} = {\pi \over 4}\,f^2h_c^2(f) = \int_0^{\infty} 
{dz\, \frac{dn}{dz}
{1\over{1+z}}~ \left.{{dE_{\rm gw}} \over {d\ln{f_r}}}\right|_{f_r = f (1+z)}}\,,
\label{pinnei}
\end{equation}
%
%
where the factor $1/(1+z)$ accounts for the redshift of gravitons and $h_c(f)$ is the characteristic GW amplitude in a frequency interval of width $\sim f$. The interpretation of equation~(\ref{pinnei}) is straightforward: the energy density per logarithmic frequency interval is equal to the integral over the cosmic history of the comoving number density of the sources, multiplied by the redshifted energy emitted by each source in the corresponding redshifted frequency range.

This general result can be specialised to the scenario considered in this paper: a stochastic background generated by a cosmic population of MBHBs during their slow adiabatic in-spiral phase. As we consider radiation generated when the sources are far from the last stable orbit, it is sufficient to model gravitational radiation at the leading Newtonian quadrupole order. Under this assumption, for a source characterised by a chirp mass $\Mc = \mu^{3/5}M^{2/5}$ -- here $\mu$ and $M$ are the reduced and total mass, respectively -- at orbital frequency $f_{\rm orb}=f_r/2$, the energy emitted per logarithmic frequency interval is (Thorne 1987)

\begin{equation}
\frac{dE_{\rm gw}}{d\ln{f_r}}=\frac{\pi^{2/3}}{3}{\cal M}^{5/3}f_r^{2/3}\,.
\label{dedlnf}
\end{equation}
Such energy is radiated between a minimum frequency, which we can safely assume to be smaller than the lowest frequency $\sim 1$ nHz probed by PTAs, and the frequency corresponding to the last stable orbit $f_\mathrm{LSO} \approx 4\times 10^{-6}$ $(1+z)^{-1} (M/10^9\Ms)^{-1}$ Hz, which is larger than the highest frequency reached in PTA observations. As a consequence, we will ignore such limits in the rest of the paper. 

The GW energy and the number density of events (i.e. mergers) throughout cosmic history depend on the MBHB masses; we therefore define
\be
\frac{dn}{dz} = \int_0^\infty d\Mc  \frac{d^2n}{dzd{\cal M}}\,,
\label{e:dndz}
\ee
where ${d^2n}/{dzd{\cal M}}$ is the comoving number density per unit redshift and chirp mass. Using equation~(\ref{e:dndz}), we can rearrange equation~(\ref{pinnei}) and obtain
\begin{equation}
h_c^2(f) =\frac{4}{\pi f^2}\int_0^{\infty} 
{dz\int_0^{\infty}d{\cal M} \, \frac{d^2n}{dzd{\cal M}}
{1\over{1+z}}~{{dE_{\rm gw}({\cal M})} \over {d\ln{f_r}}}}\,.
\label{hcdE}
\end{equation}
The previous equation provides the characteristic amplitude as a function of the number density of mergers, and it allows 
us to compute the GW spectrum using equations~(\ref{e:Omega}) and~(\ref{pinnei}).

It is also convenient to express the previous result as a function of the number of sources emitting at a given frequency. We first express the comoving number density of coalescences $dn/dz$ as a function of $d^3N/dzd{\cal M} d{\rm ln}f_r$, the comoving number of binaries emitting in a given logarithmic frequency interval with chirp mass and redshift in the range $[{\cal M},{\cal M}+d{\cal M}]$ and $[z, z+dz]$, respectively:
\begin{equation}
\frac{d^2n}{dzd{\cal M}}=\frac{d^3N}{dzd{\cal M} d{\rm ln}f_r}\frac{d{\rm ln}f_r}
{dt_r}\frac{dt_r}{dz}\frac{dz}{dV_c}\,.
\label{d3N}
\end{equation}
In the previous expression $dV_c$ is the comoving volume shell lying between $z$ and $z+dz$, and $t_r$ is time as measured in the source rest frame.
A binary slowly inspirals by losing energy and angular momentum through GW emission; assuming circular orbits gravitational radiation is emitted at twice the orbital frequency  
-- this is the dominating quadrupole order contribution to the total GW flux -- whose sky and polarisation averaged strain amplitude is given by (Thorne 1987):
\begin{equation}
h={8\pi^{2/3}\over 10^{1/2}}{{\cal M}^{5/3}\over d_L(z)}f_r^{2/3}\,,
\label{eqthorne}
\end{equation}
where $d_L(z)$ is the luminosity distance to the source. The rate of change of the emission frequency is
\begin{equation}
\frac{df_r}{dt_r} = \frac{96}{5}\,\pi^{8/3} \Mc^{5/3}f_r^{11/3}\,.
\label{e:fdot}
\end{equation}
Combining equation (\ref{dedlnf}), (\ref{eqthorne}) and (\ref{e:fdot}), we obtain
\begin{equation}
\frac{dE_{\rm gw}}{d{\rm ln}f_r}=\frac{dt_r}{d{\rm ln}f_r}\pi^2 d_L(z)^2 f_r^2 h^2\,.
\label{dEdlnfr}
\end{equation}
Substituting equations~(\ref{d3N}) and (\ref{dEdlnfr}) into equation~(\ref{hcdE}), it yields
\begin{equation}
h_c^2(f) =\int_0^{\infty} 
dz\int_0^{\infty}d{\cal M}\, \frac{d^3N}{dzd{\cal M} d{\rm ln}f_r}\,
h^2(f_r).
\label{hch2}
\end{equation}
The above equation is equivalent to equation~(\ref{hcdE}), and has a simple
interpretation: the observed characteristic squared amplitude of the GW
background is given by the integral over all the sources emitting 
in the frequency bin $d{\rm ln}f_r$ multiplied by the squared strain of each source.

It is straightforward to see from both equation~(\ref{hcdE}) and equation~(\ref{hch2}) that the predicted characteristic amplitude scales as $\propto f^{-2/3}$, with a normalisation that depends on the details of the MBHB population, and is usually represented as (e.g.  Jenet et al. 2006):
\begin{equation}
h_c(f) = h_\mathrm{1yr}\, \left(\frac{f}{{\rm yr}^{-1}}\right)^{-2/3}\,,
\label{hcpar}
\end{equation}
where $h_\mathrm{1yr}$ is a model dependent constant.

\section{Signal generation}

In order to predict the spectrum of the stochastic background generated by MBHBs, one needs to compute either $d^2n/dzd\Mc$ or ${d^3N}/{dzd{\cal M} d\ln f_r}$, see equations~(\ref{e:dndz}) and~(\ref{d3N}), according to a given cosmological model of MBH assembly; the characteristic strain amplitude $h_c(f)$ and spectrum $\Omega_\mathrm{gw}(f)$ can then be evaluated using equation~(\ref{hcdE}) or equation~(\ref{hch2}), and equation~(\ref{e:Omega}), respectively. In this section we consider a wide range of MBHB formation scenarios -- and compare them with models considered in earlier work, although this issue will be discussed in much greater detail in Section 6 -- and estimate the overall GW signal $h_c(f)$ produced by such a population. This forms the basis for our predictions about the strength and spectral shape of the {\em stochastic background} signal in the frequency range probed by PTAs that will be discussed in detail in Section 4. We will also highlight some subtleties in the computation of the background that have been ignored so far and have considerable impact on the prediction of the signal 
at frequencies $\simgt 10^{-8}$ Hz.

\subsection{Models of formation of massive black hole binary systems}

The GW spectrum produced by populations of MBHBs has already been computed in 
a handful of studies, 
based however on {\em specific} cosmic histories of MBH formation. Early works (Rajagopal \& Romani 1995, hereafter R\&R, and Jaffe \& Backer 2003, hereafter J\&B) derived the MBHB coalescence rate from available observational constraints on the fraction of galaxies that form kinematically close pairs in the local universe (Burkey et al. 1994, Colin Schramann \& Peimbert 1994, Carlberg et al. 2000), and then assigned to each galaxy a MBH according to a given MBH mass function.  A series of subsequent papers (Wyithe \& Loeb 2003, Sesana et al. 2004, 
Enoki et al. 2004) applied the EPS formalism (Press \& Schechter 1974, Lacey \& Cole 1993, Sheth \& Tormen 1999) to the hierarchical assembly of dark matter halos, using a range of prescriptions for the evolution of the population of MBHs residing in the halo centres. The halo hierarchy 
is followed both analytically (e.g. Wyithe \& Loeb 2003), or by means of Monte Carlo realisations of the merger 
hierarchy (e.g. Sesana et al. 2004). The 'observational' and EPS approach yield different estimates for the amplitude of the expected signal that need to be reconciled, and we will return to this point in great detail in Sections 5 and 6. It is worth noticing that, except for a qualitative discussion in Wyithe \& Loeb 2003, no comparison regarding the different predicted levels of the GW spectrum has been carried out so far, nor the reasons of these differences investigated. Moreover, several results show a significant departure from the simple power-law behavior $h_c \propto f^{-2/3}$ (R\&R and J\&B) and this point has been overlooked. Both these issues have a significant impact on the prospects of detection of a stochastic background with PTAs -- and implications for analysis strategies -- and will be discussed in detail.

One of the main goals of this paper is to provide a systematic exploration of different scenarios of MBH assembly, to quantify the range of signal strength and the consequences for observations with PTAs, and to explore the potential of PTAs to shed new light on the relevant epoch of cosmic history and the demographics of MBHs. We consider four classes of models of MBH assembly that qualitatively encompass the whole range 
of scenarios proposed so far and quantitatively covers a very broad spectrum of predictions for MBH formation (see Sesana, Volonteri \& Haardt 2007a for full details): (i) the VHM model 
(Volonteri, Haardt \& Madau 2003), (ii) the KBD model (Koushiappas, Bullock \& Dekel 2004), (iii) the BVRhf model 
(Begelman, Volonteri \& Rees 2006) and (iv) the VHMhopk model. In these scenarios, seed black holes are massive ($M\sim 10^4 \msun$) as in the case of KBD and BVRhf, or light ($M\sim 10^2 \msun$), as prescribed by VHM; seed black holes are abundant (VHM, KBD) or just a few (BVRhf). The VHMhopk model assumes essentially the same assembly history of the VHM model, but with a somewhat different accretion prescription 
(Volonteri, Salvaterra \& Haardt 2006). 
Each model is constructed tracing backwards the merger hierarchy of 220 dark matter halos in the mass range $10^{11}-10^{15}\msun$ up to $z=20$ (Volonteri, Haardt \& Madau 2003), then populating the halos with seed black holes and following their evolution to the present time. For each of the 220 halos all the coalescence events happening during the cosmic history 
are collected. The outputs are then weighted 
using the EPS halo mass function and integrated over the observable volume shell at every redshift to obtain numerically the coalescence rate of MBHBs as a function of black hole masses and redshift (see, e.g., figure 1 in Sesana et al. 2004 and figure 1 in Sesana et al. 2007). In other words, the outcome of this procedure is the numerical distribution $d^3N/dzd{\cal M}dt$.   

It is worth noticing that in all the above models the MBH population fits by construction {\it at every redshift} the relation, observed in local galaxies, connecting the central black hole mass $M_{\rm BH}$ and the 1-D stellar bulge velocity dispersion $\sigma$ (the so called $M_{\rm BH}-\sigma$ relation, e.g. Tremaine et al. 2002). Moreover they all include a 
gravitational recoil prescription appropriate for non-spinning black holes. It is now well established
that the recoil magnitude strongly depends on the spins (both magnitude and relative geometry) of the two black holes (Herrmann et al. 2007); the MBH build-up through merger and accretion typically results in highly spinning MBHs, 
that on average are affected by kicks with higher recoil velocity (e.g. Schnittman \& Buonanno 2007).
The adopted prescriptions for both the GW recoil and the MBH mass function in the local Universe are likely to influence the expected GW background. In Section 5 we 
map 
these effects and other poorly known dynamical processes into a  range of uncertainty of the strength of the stochastic background level.

\subsection{Semi-analytical approach} 

\begin{figure}
\centerline{\psfig{file=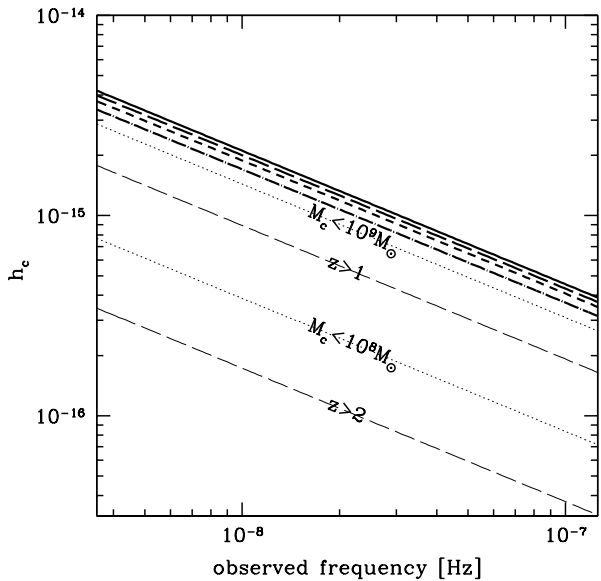,width=84.0mm}}
\caption{The characteristic amplitude $h_c$, see equation~(\ref{hch2}), 
generated by a population of massive black hole binaries 
in the frequency range accessible with Pulsar Timing Arrays 
for the assembly models discussed in the text. 
The thick lines show the total signal
for the VHM (solid line), the VHMhopk (dotted-dashed line), 
the KBD (short-dashed line) and the BVRhf (long-dashed line)
models. The thin lines show the contributions from selected portions
of the binary population predicted by the VHM model:  the contribution to the signal
considering different redshift cuts (thin dashed lines) and 
different cuts in chirp mass ${\cal M}$ (thin dotted lines).}
\label{fig1}
\end{figure}

The natural outcome of the Monte-Carlo EPS models described in the previous section consists of a list of coalescence events that are weighted on the EPS mass function and integrated in volume to obtain the numerical distribution $d^3N/dzd{\cal M}dt$ of MBHB coalescences throughout the Universe. Starting from here we easily calculate $d^3N/dzd\Mc d\ln f_r$ changing variable from $t$ to $f$ according to equation (\ref{e:fdot})  and we finally derive the characteristic amplitude $h_c$ of gravitational radiation generated from the whole population using equation~(\ref{hch2}). The very same result is equivalently obtained 
by considering the number density of coalescing binaries using equation (\ref{d3N}) and substituting equation (\ref{dedlnf}) into equation (\ref{hcdE}). This is the approach used by several authors in the past, though only for selected models. The resulting estimated characteristic amplitude $h_c$ for each of the four models VHM, VHMhopk, BVRhf and KBD are shown in Fig. \ref{fig1}. Not surprisingly,  the characteristic amplitude scales always as $h_c\propto f^{-2/3}$ -- it must be so by construction -- and the level of the signal $h_\mathrm{1 yr} \approx 10^{-15}$ is similar for all of them;
one can expect a much larger spread of values due to the uncertainty on specific model parameters, which we shall discuss in Section 5, rather than the specific prescription for MBH assembly. 

The fact that in the window $10^{-9}$ Hz - $10^{-7}$ Hz the level of the signal 
does not strongly depend on the specific assembly scenario 
is easily explainable: the major contributors to very low frequency GWs are massive (${\cal M}>10^8\msun$) and nearby ($z<2$) binary systems,
whose population is fairly well constrained by present observations. Whether these MBHs
are the end-product of light or massive seeds has only modest 
effects on the signal in the PTA frequency window. We recall however that the models considered here assume a MBH mass function based on the $M_{\rm BH}-\sigma$ relation. We will see in Section 5 that different assumptions for the MBH--stellar bulge relation have a significant (a factor of $\gsim$ 2) impact on the overall level of the signal.

\subsection{Monte-Carlo approach}

A different procedure to estimate the signal produced by a population of
MBHBs is to directly build such a population through a Monte-Carlo 
approach and compute the overall signal in the
relevant frequency range by adding the contributions from each individual 
binary, using the same merger trees. In observations with PTAs, radio-pulsars are monitored
weekly for periods of years. The relevant frequency band is therefore 
between $1/T$ -- where $T$ is the total observation time -- and the Nyquist
frequency $1/(2\Delta t)$ -- where $\Delta t$ is the time between two adjacent 
observations, corresponding to $5\times 10^{-9}$ Hz - $10^{-7}$ Hz. 
The frequency resolution bin is $1/T$, and a source can be considered 
effectively monochromatic if during the observation time the frequency 
shift $T df/dt$ (due to loss of energy through GW emission) is less than $1/T$. It
is simple to verify through equation~(\ref{e:fdot}) that this is indeed the case
for the overwhelming majority of sources in the relevant mass range for
$f \simlt 10^{-7}$ Hz.

Starting from a model-dependent prescription of the MBH assembly history,
one generates a population of MBHBs using a Monte-Carlo sampling of
the trivariate distribution $d^3N/dzd{\cal M} d{\rm ln}f_r$; the total signal $h_c$ 
is then computed by adding all the contributions in every frequency
interval $1/T$. A key difference with respect to the semi-analytical approach
described in Section 3.2 is that here we do not directly perform the integral 
in equation~(\ref{hch2}), but, given the Monte-Carlo sample of the emitting sources
we simply sum the contributions of each individual one. It is also obvious -- but
this point is essential to explain discrepancies with the semi-analytical approach --
that by default one takes into account the {\it discrete nature} of the GW emitters.
 
\begin{figure}
\centerline{\psfig{file=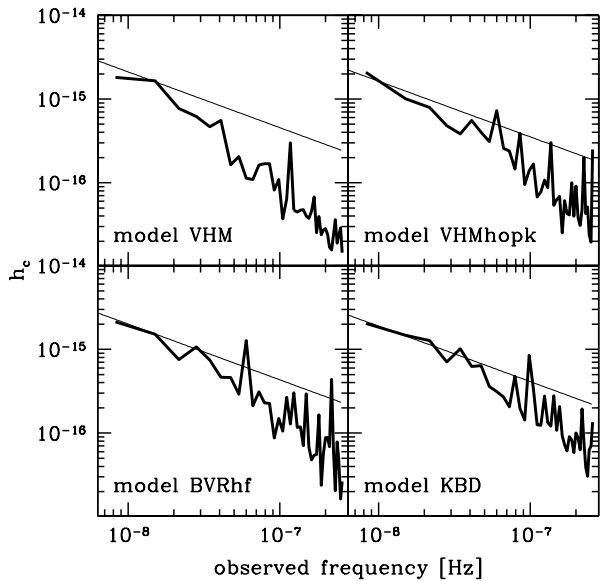,width=84.0mm}}
\caption{The total contribution to the characteristic amplitude $h_c$ of
the GW signal from a population of massive black hole binaries 
in the frequency range accessible 
to Pulsar Timing Arrays. In each panel, the thick line shows $h_c$ produced in 
a specific Monte-Carlo realization and compares it to the prediction yielded 
by the semi-analytical approach (thin line). The observation time is 
$T = 5$ yr. }
\label{fig2}
\end{figure}
\begin{figure}
\centerline{\psfig{file=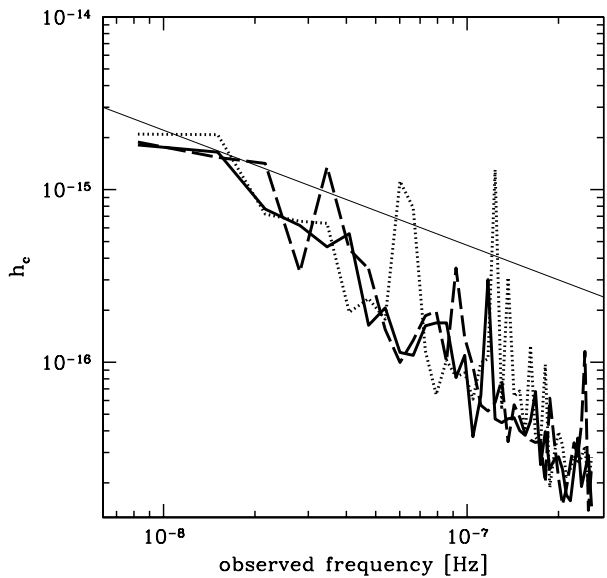,width=84.0mm}}
\caption{Same as Fig.~\ref{fig2}, but now three different Monte-Carlo 
realizations of the GW signal expected from the {\em same} massive black hole
assembly model (VHM) are 
shown as a function of frequency (thick solid, dotted and dashed line). 
The thin solid line is the prediction obtained using the semi-analytical approach for 
the same model. The observation time is set to 5 years.}
\label{figmore}
\end{figure}

The results of this approach are reported in Fig. \ref{fig2} for the different models of
MBH assembly discussed in Section 3.1. Each panel shows $h_c$ obtained
through both the semi-analytical approach and the Monte-Carlo realization in 
a frequency resolution bin $\Delta f=1/T$, where $T$ is set to 5 years. In Fig.
\ref{figmore} we focus on a specific MBH assembly model (VHM) and compare
three different Monte-Carlo realizations of the GW signal. Some distinctive
features are immediately clear: (i) the Monte-Carlo and semi-analytical approach
provide estimates of the total GW signal that are very well in agreement at frequencies
below $\approx 10^{-8}$ Hz; (ii) at frequencies $\simgt 10^{-8}$ Hz the total contribution
to $h_c$ produced through the Monte-Carlo procedure falls clearly below the
semi-analytical prediction $h_c \propto f^{-2/3}$ suggesting a steepening of
the spectrum in this frequency range. 
We would like to highlight that such behavior is also present in the results reported
in earlier works (e.g. R\&R and J\&B) that employ an analogous Monte-Carlo approach,
but consider different assembly scenarios. Such discrepancy has never been discussed so far 
but is actually symptomatic of an important conceptual difference between the
two approaches, with direct observational implications. We will devote the
next section to investigate and explain this difference in detail. Before proceeding,
we would also like to stress that so far we have referred to $h_c$ as the 
characteristic amplitude of the \emph{total GW signal} rather than the \emph{GW stochastic background}
from a MBHB population; once more this is a crucial difference that will become
clear in the next section. 

\section{Stochastic background}
\label{s:small}

In this section we investigate the origin of the discrepancy between the semi-analytical and Monte-Carlo estimate of the total signal produced by a population of MBHBs and provide a new estimate for the characteristic amplitude, or equivalently the spectral content of the relevant {\it stochastic background} valid over the whole frequency range of PTA observations.
In Section~\ref{ss:origin} we study the details of the source population (MBHB distribution in redshift, mass and frequency) that make up the signal, and their consequences. In Section~\ref{ss:parametric} we consider a heuristic model that explains the inconsistency between the two approaches, and provide a simple analytical expression for the stochastic background as a function of three parameters. In Section~\ref{ss:toy} we present an analytical approximation to the Monte-Carlo results to further clarify the key conceptual issues; readers not interested in the mathematical details can skip this section without loss of any crucial information for the rest of the paper. We conclude by summarising the subtleties related to the computation of the overall signal and the pitfalls of the naive semi-analytical approach in Section 4.4. We concentrate on the VHM model to discuss the fundamental issues; the results are qualitatively identical for all the other astrophysical scenarios, with differences concerning only the numerical values of the relevant quantities, that are presented in Table \ref{tab2}.

\subsection{Origin of the discrepancy}
\label{ss:origin}

\begin{figure}
\centerline{\psfig{file=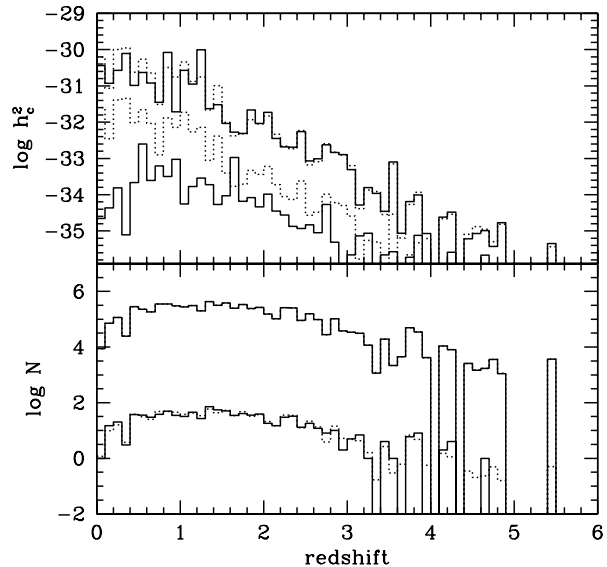,width=84.0mm}}
\caption{Contribution of different redshift intervals to the build-up of the GW signal at two different frequencies,  $f=8\times 10^{-9}$ Hz and $10^{-7}$ Hz, as computed using the two different approaches described in the text: Monte-Carlo sampling (solid lines) and semi-analytical approach (dotted lines). In each panel the upper histograms refer to $f=8\times 10^{-9}$ Hz and the lower histograms refer to $f=10^{-7}$ Hz. These results refer to the VHM model.}
\label{fig3}
\end{figure}

\begin{figure}
\centerline{\psfig{file=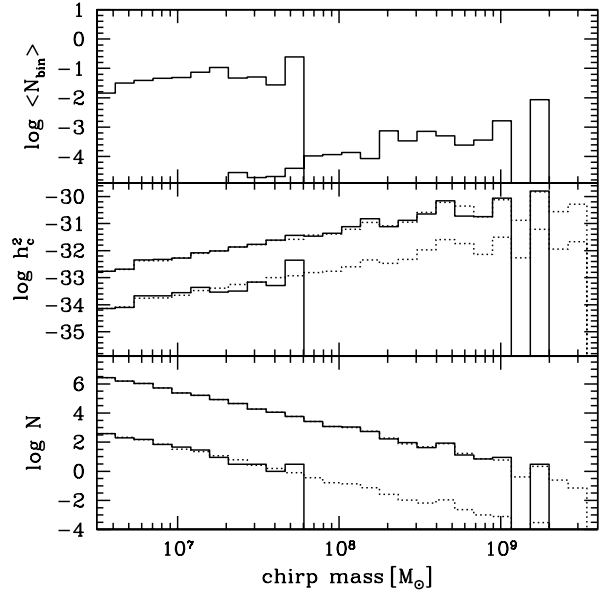,width=84.0mm}}
\caption{Same as Fig. \ref{fig3}, but now considering the contribution from different intervals in chirp mass. In this case we also show (top panel) the mean number of frequency bins $\langle N_\mathrm{bin}\rangle$  spanned by a binary during 5 years, as a function of ${\cal M}$. The line-style is the same as in Fig.~\ref{fig3}.}
\label{fig4}
\end{figure}

We start by studying the contributions to $h_c(f)$ at different frequencies as a function of $z$ and ${\cal M}$. We consider the contributions to the signal centered at the two fiducial frequencies $f = 8\times 10^{-9}$ Hz (close to the minimum accessible frequency for a $T=5$ yr observation) and $10^{-7}$ Hz, with width $\Delta f = 1/T \simeq 6\times 10^{-9}$ Hz, from sources within the shell $z$ and $z+\Delta z$, where $\Delta z = 0.05$. Fig.~\ref{fig3} clearly shows that the Monte-Carlo and semi-analytical approach provide the same estimate of the total number of sources per redshift interval that contribute to the signal. The estimate of the signal $h_c$ is again consistent between the two approaches at $f = 8\times 10^{-9}$ Hz; however, at  $f = 10^{-7}$ Hz the estimate of $h_c$ inferred using the Monte-Carlo sampling of the population is significantly smaller than that computed semi-analytically. This behaviour can be understood by studying the contribution to the signal by the binaries in the population as a function of ${\cal M}$; this is shown in Fig. \ref{fig4}. We first see that the average number $\m N_\mathrm{bin}\M$ of frequency resolution bins of width $1/T$  spanned by each binary during the observation is always smaller than one; this ensures that we can safely treat all the binaries contributing to the background as stationary sources. Further, notice that the main contribution to the characteristic amplitude $h_c$ is generated by the high-mass tail of the ${\cal M}$ distribution. At frequencies $\simlt 10^{-8}$ (upper curves in each panel of Fig. \ref{fig4}) the number of sources contributing to the signal is $\sim 10^2-10^6$ (depending on the mass range) and the Monte-Carlo and semi-analytical approach yield the same estimate for $h_c$. However, at frequencies $\simgt$ a few $\times 10^{-8}$ Hz (bottom curves in each panel of Fig. \ref{fig4}) the overwhelming majority of the radiation inferred using the semi-analytical approach is actually produced by less than one source with ${\cal M} \simgt 10^8\,\Ms$. This is clearly an artefact and the key point to understand the difference of the results: a Monte-Carlo sampling of the population returns correctly a discrete number of sources -- in this case either one or none -- and not fractions of a binary; therefore it cannot possibly (and it should not!) reproduce the semi-analytical result, that contains spurious contributions from "fractional sources" that are clearly not present; the actual signal is significantly smaller than the one computed semi-analytically. 

\begin{figure}
\centerline{\psfig{file=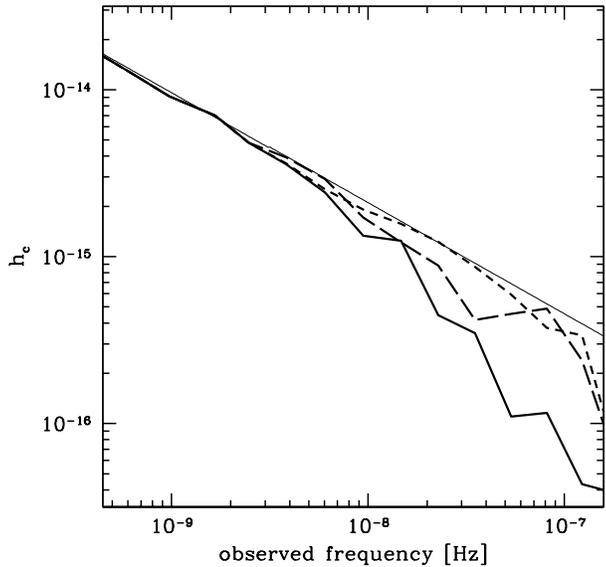,width=84.0mm}}
\caption{The very low frequency characteristic amplitude $h_c$ of
the GW signal predicted by the VHM model.
The thin and thick solid lines show $h_c$ computed
using the semi-analytical and Monte-Carlo approach, respectively. The 
thick long--dashed and short--dashed lines shows the signal from
the Monte-Carlo realizations obtained by {\em artificially} multiplying the 
number of the contributing sources by a factor 10 and 100, respectively,
and dividing the contribution of each sources to $h_c$ by the same factor.}
\label{small}
\end{figure}

To cross-check the discrepancy of the two results, we can {\it artificially} increase the number of sources present in the sample, but divide by the appropriate factor the contribution of each binary to $h_c$. The semi-analytical estimate is clearly unaffected and produces the familiar $h_c \propto f^{-2/3}$ dependency over the whole frequency band; however, as the number of sources increases, the Monte-Carlo result matches progressively better across the entire frequency range the previous result. This is reported in Fig.~\ref{small}: the number of sources in (a given model of) the Universe is fixed, and the actual total GW signal is far from resembling the simple $f^{-2/3}$ power law over the whole frequency range.


The semi-analytical approach, that treats the distributions of sources as {\it continuous}, leads to a significant over-estimate of the strength of the background in an important frequency range for PTA observations. In order to derive a correct prediction of the level of the GW stochastic backgrounds it is essential that the \emph{discrete} nature of the sources is consistently taken into account. To reiterate this point, when one evaluates the coalescence rate of MBHBs (e.g. using merger tree models), one performs different realisations of the assembly of individual halos, and then weights each halo realisation over the EPS halo mass function, that is $\ll 1$. As a consequence, the rate of massive coalescing binaries at low redshift could be, for example $\sim 10^{-5}$ yr$^{-1}$. When this rate is used to compute the number of binaries contributing to a certain frequency interval  (see equation (\ref{d3N})), one finds a number of sources $\ll 1$ for $f \simgt 10^{-8}$.  As a matter of fact, one performs an idealised average  realisation of the Universe. In practice the same average is done when extrapolating the coalescence rate from the observation of the fraction of close galaxy pairs. Once more, by dividing the number of close pairs by the typical merging time, one obtains rates that are $\ll 1$ yr$^{-1}$; the associated number of emitting sources is again $\ll 1$ (and for consistency, it could not be otherwise). To summarize, the discrete nature of sources is critical; semi-analytical evaluations of the background (that treat each distribution function as continuous) leads to spurious predictions in the observational window above $\sim 10^{-8}$ Hz.

\subsection{Phenomenological model}
\label{ss:parametric}

It is useful to provide a parametrised expression of $h_c$ to replace the incorrect (in parts of the frequency spectrum) simple dependency 
$h_c \propto f^{-2/3}$.
We start by providing a simple way to quantify the spurious additional contribution to $h_c$ given by semi-analytical approaches.

\begin{figure}
\centerline{\psfig{file=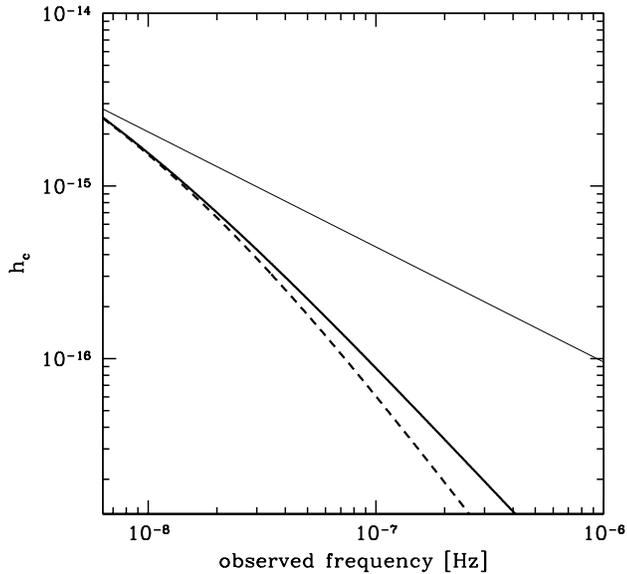,width=84.0mm}}
\caption{Signal spectrum in the VHM model. The characteristic strain is plotted versus the observed frequency. The thin solid line shows the na\"{\i}ve  semi-analytical estimate of the amplitude from equation (\ref{hcdE}) or (\ref{hch2}), the thick solid line is typical level of the 'real' signal once the exceeding spurious contribution is subtracted according to equation (\ref{missing}), and the thick dashed line represents the level at which the signal can be considered a stochastic background (assuming a 5 years observation) according to the subtraction given by equation~(\ref{stochastic}).}
\label{figfit}
\end{figure}

Let us consider a frequency $\tilde{f}$, and integrate the total number of sources emitting at $f>\tilde{f}$ in a given mass range $[\tilde{\cal M}, +\infty[$. For any given $\tilde{f}$, we can always find a $\tilde{\cal M}$ such that
\be
\int_{\tilde{\cal M}}^{\infty}  d{\cal M} \int_{\tilde{f}}^{f_{\rm LSO}}  df\, \frac{d^2N}{d{\cal M} df}=1\,.
\label{missing}
\ee
This means that all the signal due to sources with ${\cal M}>\tilde{\cal M}$
at $f>\tilde{f}$ is generated by less than one source, and it is therefore not
present in the radiation produced by an actual population of MBHBs. An example is illustrated in Fig.
\ref{figfit}. If we consider for example $\tilde{f}=10^{-7}$ Hz, the implicit
solution of equation (\ref{missing}) is $\tilde{\cal M} \simeq 1.5\times 10^8\,\Ms$;
GWs due to MBHBs with heavier chirp mass are generated by less than one source, and so they do not contribute to the signal.

\begin{figure}
\centerline{\psfig{file=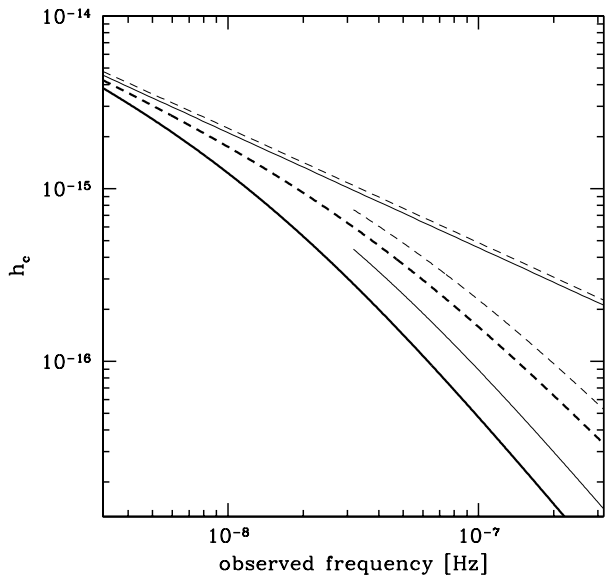,width=84.0mm}}
\caption{Fits to the stochastic background assuming 1 (thin lines) and 10 (thick lines) years of observations. The $f^{-2/3}$ straight lines represent the na\"{\i}ve semi-analytical background predictions. Solid lines are for the VHM model, while dashed lines are for the KBD model.}
\label{figfits}
\end{figure}

What we have computed so far is the {\it total contribution} to the characteristic amplitude $h_c(f)$ as a function of frequency from the whole population of MBHBs in the Universe (for different models). However, here we are interested in estimating the {\it stochastic background} from the population. Whether the superposition of many deterministic signals should be effectively considered a stochastic signal depends on the frequency range and the observation time. Here we will consider the usual (simplified) criterion that the signal is stochastic if the number of sources whose radiation contributes to the frequency bin of width $1/T$ centered at $f$ is $\gg 1$; the stochastic level of the signal is then the amplitude of the sum of the individual contributions.
Note therefore that Fig.~\ref{fig2} and Fig.~\ref{figmore} do not show the GW background contribution, but the total signal from the whole population of MBHBs. To evaluate the GW stochastic contribution, we can follow the same approach that has led us to equation~(\ref{missing}), by simply replacing the range of integration over frequency with $[\tilde{f},\tilde{f} + 1/T]$; we can therefore find a value $\tilde{\cal M}$ such that 
\be
\int_{\tilde{\cal M}}^{\infty} d{\cal M} \int_{\tilde{f}}^{\tilde{f}+1/T} df\, \frac{d^2N}{d{\cal M} df}=1\,.
\label{stochastic}
\ee
The integral above identifies the sources in the population that do not contribute to the stochastic background. The result depends on the observational time, since the longer the observation, the narrower the frequency bin and, accordingly, the lower the level of signal contribution that can be considered stochastic. We note however that the ${\cal M}$-distribution of sources is typically a steep power law (cf. Fig. \ref{fig4}). If we halve the frequency bin by doubling $T$, $\tilde{\cal M}$ in equation (\ref{stochastic}) would be only slightly reduced, and the level of stochasticity would change by less than $30\% $. We find that the stochastic background does change by a factor $\simlt 2$ for observational times 1 yr$\simlt T \simlt $ 10 yrs, as shown in Fig. \ref{figfits}. Moreover, for typical observation times of several years, the number of sources radiating in the frequency interval $[\tilde{f}$,$\tilde{f}+1/T]$ is of the same order of those radiating in the interval $[\tilde{f}$,$f_{\rm LSO}]$ (this is a consequence of equation (\ref{e:fdot}), that implies that the number of sources radiating at a given frequency $f$ is proportional to $f^{-11/3}$); as a consequence the amplitude of the stochastic signal is only slightly lower than the one estimated following the condition given in equation (\ref{missing}). Fig. \ref{figfit} shows this effect; this result is consistent with the outcome of the study of the total number of sources that contribute to the total radiation from a population of MBHBs (at selected frequencies and in a range of either redshift or chirp mass) that is shown in the bottom panel of Fig. \ref{fig3} and Fig. \ref{fig4}.
  

\begin{table}
\begin{center}
\begin{tabular}{cccc}
\hline
MODEL & $h_0$ & $f_0$ & $\gamma$\\
              & ($\times 10^{-15}$) & ($\times 10^{-8}\,\mathrm{Hz}$) & \\
\hline
   VHM&       2.15 & 1.42 & -1.09 \\
   VHMhopk&   0.69 & 4.27 & -1.08 \\
   KBD&       0.67 & 5.24 & -1.04 \\
   BVRhf&     0.89 & 3.95 & -1.11 \\
\hline
\end{tabular}
\end{center}
\caption{Value of the parameters for the
analytical expression 
of the characteristic amplitude of a GW stochastic background
given by equation (\ref{hfit}).}
\label{tab2}
\end{table}

Using the ${\cal M}$-distributions predicted by the merger trees (as those 
shown in the lower panel of Fig. \ref{fig4}),
we have calculated $\tilde{\cal M}$ as a function
of $\tilde{f}$ according to both equations (\ref{missing}) and (\ref{stochastic}).
Then we have subtracted at every frequency the portion of the signal due to the 
sources with ${\cal M}>\tilde{\cal M}$. Best fits to the residual
signals obtained in this way are plotted in Fig. \ref{figfit} for the 
VHM model; all the other models show a similar behaviour.
The signal deviates significantly from a simple $f^{-2/3}$ power-law
for $f\simgt 10^{-8}$ Hz. The GW stochastic background is well fitted 
by a function of the form
\be
h_c(f)=h_0\left(\frac{f}{f_0}\right)^{-2/3}\left(1+\frac{f}{f_0}\right)^{\gamma}\,.
\label{hfit}
\ee 
The three parameters $h_0$, $f_0$ and $\gamma$ are given in Table \ref{tab2}
for the four fiducial MBHB assembly scenarios considered in this paper.
In general, the slope of the stochastic contribution to $h_c$ starts to deviate from $-2/3$ at around $10^{-8}$ Hz, and becomes as steep as $\approx -1.5$.
In the remainder of the paper, we will use equation~(\ref{hfit}) and Table~\ref{tab2} as estimates of the characteristic amplitude of the GW stochastic background produced by MBHB populations. Astrophysical uncertainties affecting the fit parameters will be discussed in Section 5, and estimated ranges for $h_0$, $f_0$ and $\gamma$ are given in equations~(\ref{fitparameters1}-\ref{fitparameters3}). In the next sub-section we provide a different derivation of the main results that we have just presented. 

\subsection{The discrepancy revisited}
\label{ss:toy}

The main result reported in equation~(\ref{hfit}) has been obtained by carrying out Monte-Carlo sampling of the assembly history of MBHBs and then by fitting the numerical result. 
It is however instructive to derive the same result using a simple analytical model that relies only on general assumptions: the mass function of black holes, the GW quadrupole formula and the discreteness of sources. 

Let us consider the probability density function of sources per unit chirp mass  $p({\cal M})$, so that
\be
\int_0^{\infty}p({\cal M})\,d{\cal M}=1\,.
\label{probdens}
\ee
We sample now $p({\cal M})$ with $N$ objects (sources). The total number of objects $N$ is model dependent, but we keep it general for the time being; we can define $\tilde{{\cal M}}(N)$ such that
\be
\int_{\tilde{{\cal M}}}^{\infty}p({\cal M})\,d{\cal M}=\frac{1}{N}\,.
\label{sampling}
\ee
For any function $h$ -- this will be the GW strain -- we can always define its value weighted by the probability distribution $p({\cal M})$ as 
\be
{\bar h}^2=\int_0^{\infty}p({\cal M})h^2({\cal M})\,d{{\cal M}}\,.
\ee
For a fixed value of $N$, we define the following quantity:
\be
Z\equiv\frac{\int_{\tilde{{\cal M}}}^{\infty}p({\cal M})h^2({\cal M})\,d{\cal M}}{\int_0^{\infty}p({\cal M})h^2({\cal M})\,d{\cal M}}\,;
\label{fraclost}
\ee
$Z$ is the fraction of $h^2$ whose contribution comes from less than one source, and is therefore not actually present. The overall value of $h$ sampled with $N$ objects is then simply
\be
h_{\rm eff}= {\bar h}\sqrt{1-Z}\,.
\label{heff}
\ee

We can apply this general framework, and in particular equation~(\ref{heff}) to the problem of estimating the effective stochastic background from a population of sources. To do this we need to evaluate, at any given frequency $f$, the probability density function per unit chirp mass and unit frequency $p({\cal M},f)$ and the number of sources $N(f)$ given by the distribution of sources $d^2N/d{\cal M} df$ emitting in a frequency interval of width $\Delta f = 1/T$ centered on $f$. Firstly we note that $p({\cal M},f)=p({\cal M})$, i.e. the probability distribution does not depend on $f$. Ignoring subtleties related to the redshift we can indeed write
\ba
\frac{d^2N}{d{\cal M} df} & = & \frac{d^2N}{d{\cal M} dt}\frac{dt}{df}
\nonumber\\
& =& \frac{5\Mc^{-5/3}}{96\pi^{8/3}}\frac{d^2N}{d{\cal M} dt} f^{-11/3},
\ea
where the second equality comes from equation~(\ref{e:fdot}); the merger rate $d^2N/d{\cal M} dt$ is independent of $f$. The frequency dependence factorises and we obtain 
\ba
p({\cal M},f) & = & \frac{\int_{f-\Delta f/2}^{f+\Delta f/2} df'\, \frac{d^2N}{d{\cal M} df'}}{\int_{0}^{\infty} d{\cal M} \int_{f-\Delta f/2}^{f+\Delta f/2} df'\, \frac{d^2N}{d{\cal M} df'}}
\nonumber\\
& = & \frac{\Mc^{-5/3}\frac{d^2N}{d{\cal M} dt}}{\int_{0}^{\infty} \Mc^{-5/3}\frac{d^2N}{d{\cal M} dt}d{\cal M}} = p({\cal M})\,.
\label{prob}
\ea
In most of the models of MBH assembly (e.g. Volonteri Haardt \& Madau 2003), $d^2N/d{\cal M} dt$ is a function well approximated by a single power law $\propto {\cal M}^{-\beta}$ with $1.5\, \lsim\, \beta\, \lsim\, 2$; the chirp mass probability density function, equation (\ref{prob}), is therefore well described by {the following steep power law
\be
p({\cal M}) =
\left\{
\begin{array}{ll}
A{\cal M}^{-\alpha}
&  \quad \quad {\cal M}_\mathrm{min} \le {\cal M} \le {\cal M}_\mathrm{max}
\\
0 
&  \quad \quad \mathrm{elsewhere}
\end{array}\,,
\right.
\label{mfunction}
\ee
where $3 \,\lsim\, \alpha \,\lsim\, 3.5$, ${\cal M}_{\rm min}$ is of the order of the first MBH seed mass ($\sim 10^{2}\,\Ms$ for the VHM and the VHMhopk models; $\sim10^{4}\,\Ms$ for the BVRhf and the KBD models) and ${\cal M}_{\rm max}\simeq 1-3\times 10^9\msun$. The normalisation constant $A$ is set by equation~(\ref{probdens}). The quantity $N(f)$ on the right-hand side of equation~(\ref{sampling}) however depends on frequency. Using the quadrupole formula, see equation~(\ref{e:fdot}), the number of sources per frequency bin $\Delta f = 1/T$ is given by 
\be
N\left(f,\Delta f\right)=N_0\left(\frac{f}{f_0}\right)^{-11/3}
\label{ndif}
\ee
where $f_0$ is a reference frequency, and $N_0 = N\left(f_0,\Delta f\right)$ is the number of sources emitting in the frequency bin $\Delta f$ centred on $f_0$, namely 
\be
N_0 =\int_0^{\infty}d{\cal M}\int_{f_0-\Delta f/2}^{f_0+\Delta f/2}df
\frac{d^2N}{d{\cal M} df},
\ee
that depends on the specific MBH assembly scenario. Substituting equations (\ref{mfunction}) and (\ref{ndif}) into equation (\ref{sampling}) we find
\be
A\int_{\tilde{{\cal M}}(f)}^{{\cal M}_{\rm max}}{\cal M}^{-\alpha}\,d{\cal M}=\frac{1}{N_0}\left(\frac{f}{f_0}\right)^{11/3}\,.
\label{e:A}
\ee
Solving the previous equation for $\tilde{{\cal M}}$ yields:
\be
\tilde{{\cal M}}(f)={\cal M}_{\rm max}\left[1+\frac{\alpha-1}{A\,N_0{\cal M}_{\rm max}^{1-\alpha}}\left(\frac{f}{f_0}\right)^{11/3}\right]^{1/(1-\alpha)}.
\label{mbar}
\ee
The previous equation is clearly valid only if $\alpha \ne 1$, which is indeed the case considered here (cf. the discussion following equation~\ref{mfunction}).
Using the fact that $h_c^2 \propto {\cal M}^{10/3}$ and substituting $p({\cal M})$ from equation~(\ref{mfunction}) into equation~(\ref{fraclost}), the fraction of signal that is not present due to the discrete nature of sources is
\begin{eqnarray}
Z(f) & = & \frac{\int_{\tilde{{\cal M}}(f)}^{{\cal M}_{\rm max}}{\cal M}^{10/3-\alpha}\,d{\cal M}}
{\int_{{\cal M}_{\rm min}}^{{\cal M}_{\rm max}}{\cal M}^{10/3-\alpha}\,d{\cal M}}
\nonumber\\
& \simeq & 1-\left(\frac{\tilde{{\cal M}}(f)}{{\cal M}_{\rm max}}\right)^{13/3-\alpha}\,.\label{zeta}
\end{eqnarray}
Here we have used the fact that for GWs from MBHBs $\tilde{{\cal M}}(f) \gg {\cal M}_{\rm min}$. The value of $\tilde{{\cal M}}(f)$ depends on the frequency. Looking at equation~(\ref{mbar}), one can identify a critical value of the frequency $\bar{f}$ -- for the merger trees used in this paper $\bar{f} \sim10^{-8}$ Hz -- that separates two different regimes. In the frequency region $f\simlt \bar{f}$, the signal is characterised by}  $\tilde{{\cal M}} \approx {\cal M}_{\rm max}$ and therefore $Z(f) \ll 1$: effectively the overwhelming majority of the sources 
equally contribute to the stochastic background.
On the other hand, for $f>\bar{f}$ we obtain
\be
\tilde{{\cal M}}(f) \sim {{\cal M}}_{\rm max}\left(\frac{f}{\bar{f}}\right)^{11/(3-3\alpha)}\,\ll{{\cal M}}_{\rm max}\,;
\label{mtilde}
\ee
this implies that at higher frequencies $Z(f) \approx 1$ and the vast majority of the signal is due to less than one source, that clearly has no physical meaning; indeed Monte-Carlo realisations of the background do not contain such a contribution. If we now insert equation~(\ref{mtilde}) into equation~(\ref{zeta}) and use equation~(\ref{heff}), we find that the level of stochastic background is
\be
h_{\rm eff}(f)= \bar{h}(f)\sqrt{1-Z(f)}\propto f^{-2/3-11/[6(1-\alpha)]}\,;
\label{heffdep}
\ee
%
for $3\, \lsim\, \alpha\, \lsim\, 3.5$, the previous expression yields $h_{\rm eff}(f) \propto f^{-1.6}-f^{-1.4}$, which is in good agreement with equation~(\ref{hfit}) for $h_c(f)$, that was obtained by fitting the Monte-Carlo realisations. 

The above result depends on the width of the frequency bin because it describes the level at which the signal can be considered stochastic. However the discrepancy between the overall signal from a MBHB population computed using the semi-analytical and the Monte-Carlo approach that we have discussed in the previous sections and is summarised in equation (\ref{missing}) depends only on the details of the specific population and not on the observational time. We can easily derive this result using the mathematical framework developed in this section; we therefore compute the effective strength of the total signal by simply considering, at any given frequency, all the sources emitting at that frequency or above. The probability distribution given by equation (\ref{mfunction}) does not change, but equation (\ref{ndif}) needs to be replaced by  
\be
N(f)={N}_0'\left(\frac{f}{f_0}\right)^{-8/3}\,,
\label{ndif2}
\ee
where 
\be
N_0'=\int_0^{{\cal M}_{\rm max}}d{\cal M}\int_{f_0}^{f_{\rm LSO}}df \frac{d^2N}{d{\cal M} df}.
\label{normaliz2}
\ee
Using equation (\ref{ndif2}), we are considering as sampling sources all the binaries contributing to the signal above the frequency $f$, and the formalism becomes independent of the frequency bin. Replacing equation (\ref{ndif}) with equation (\ref{ndif2}) and re-evaluating equations (\ref{e:A})-(\ref{zeta}),
the quantity $Z(f)$ is now interpreted as the spurious total signal excess at the frequency $f$, according to equation (\ref{missing}). In this case $\tilde{\cal M}\propto f^{8/(3-3\alpha)}$ for $f>\bar{f}$. Inserting this dependence into equation  (\ref{zeta}) and substituting into equation (\ref{heff}) we find the actual total signal level to be $h_{\rm eff}(f)\propto f^{-1.35}-f^{-1.2}$, for $3\,\lsim\,\alpha\,\lsim\,3.5$.

\subsection{Summary of the results}

\begin{figure*}
\centerline{\psfig{file=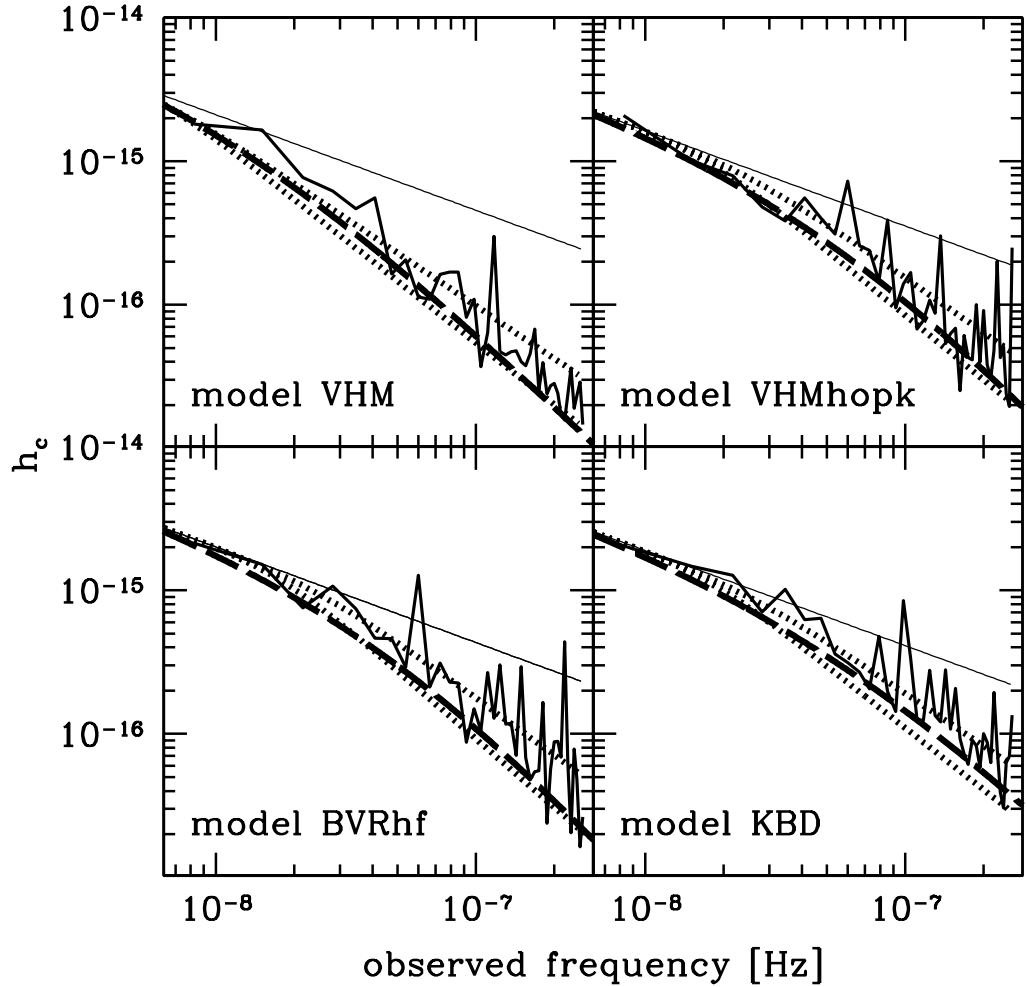,width=144.0mm}}
\caption{
The very-low frequency GW signal for the four assembly scenarios explored in the paper. In each panel, the thin line represents the na\"{\i}ve standard semi-analytical amplitude $h_c\propto f^{-2/3}$ while the spiky solid line shows the output of a specific Monte-Carlo realisation of the radiation from the whole MBHB population assuming a 5 yrs observation. The lower dotted line marks the level at which there is at least one source per frequency bin ($\Delta f\sim 6\times10^{-9}$) according to the model developed in Section \ref{ss:toy}, and the upper dotted line shows the signal amplitude from the whole population, again evaluated following the model discussed in Section \ref{ss:toy}.  
The dashed line is the best fit to the Monte-Carlo realisation of the stochastic background given by equation  (\ref{hfit}) and Table ~\ref{tab2}.}
\label{figdrop}
\end{figure*}

A summary of the results is shown in Fig. \ref{figdrop} for all the four reference models. As expected, the two curves representing the level of the actual total signal and the stochastic background are quite similar in the frequency range of interest; this is due to the fact that $dN/df\propto f^{-11/3}$. In fact, as we have discussed in Section \ref{ss:parametric}, the number of sources in a frequency bin around $f$ is of the order of the number of sources emitting in the interval $[f,+\infty[$. The slopes of the curves lie in the expected range: $f^{-1.6}-f^{-1.4}$ for the stochastic background, and $f^{-1.4}-f^{-1.2}$ for the total signal. The agreement among the Monte-Carlo simulated background, the fitting formula given by equation~(\ref{hfit}) with parameter values shown in Table ~\ref{tab2}, and the outcome of the model presented in Section \ref{ss:toy} is good. In conclusion, the expected signal consists of a stochastic background well described by a superposition of two power laws (equation (\ref{hfit})), on top of which individual outliers are present due to rare, nearby bright sources. The slopes of the power laws are $-2/3$ at low frequencies, and $\approx -1.5$ at high frequencies, with a knee around $\sim 10^{-8}$ Hz. Both the knee frequency and the index of the slope at high frequency depend on the details of the ${\cal M}$-distribution predicted by the MBH assembly model. For $f>10^{-8}$ Hz, most of the signal predicted by semi-analytical approaches is actually not present because of the {\it discrete nature of sources}.

\section{Sources of uncertainty}

We have shown that the stochastic background is well described by equation (\ref{hfit}), that
depends on the three parameters $h_0$, $f_0$ and $\gamma$, see Table~\ref{tab2}. The amplitude 
$h_0$ sets the overall strength of the signal; equation (\ref{hch2}) shows that
$h_c^2(f)$ is proportional to the
number $N$ of sources emitting at frequency $f$ multiplied by the contribution $h^2$ of each individual source. Since $h^2\propto{\cal M}^{10/3}$, we have:
\begin{equation}
h_0\propto \sqrt{N}\,{\cal M}^{5/3}\,. 
\label{dependences}
\end{equation}
The overall level of the GW stochastic background is therefore set by $N$, that depends on the merger rate of 
galactic halos, the MBH occupation fraction and the efficiency of the 
MBHB coalescence, and by ${\cal M}$, that depends on the hosted MBH mass function and on the typical mass ratio of the coalescing binaries. There are considerable uncertainties surrounding the values of these parameters and others that enter the relevant physical processes that ultimately affect the amplitude of the GW stochastic background. In this section we {\em quantify} such uncertainties according to our current theoretical understanding and observational data, and, as a consequence the prospects of detection with present and future PTA surveys. 

Assuming that  $N$ and ${\cal M}$ are uncorrelated, uncertainties in $N$ and ${\cal M}$ will be reflected into a range of possible values for $h_0$; we indicate with $\sigma_{N}$ and $\sigma_{\cal M}$ the root-mean-square (rms) spread of values in ${N}$ and ${\cal M}$, respectively, computed from sampling a number of possible models and scenarios; this translates into a rms range of possible levels of the GW stochastic background, denoted by $\sigma_{h_0}$ around a mean value $\langle h_0 \rangle$. The model uncertainties $\sigma_{N}$, $\sigma_{\cal M}$ and  $\sigma_{h_0}$ are related through equation ~(\ref{dependences}) by:
\begin{equation}
\left(\frac{\sigma_{h_0}}{h_0}\right)^2=\left(\frac{1}{2}\frac{\sigma_{N}}{N}\right)^2+\left(\frac{5}{3}\frac{\sigma_{\cal M}}{{\cal M}}\right)^2\,.
\label{error}
\end{equation}
A rigorous quantitative estimate of the impact of uncertainties in 
the merger rate and the 
MBH mass function on $f_0$ and $\gamma$ is not 
straightforward. We note however that for the four models considered in this paper $\gamma$ 
varies only by $\approx$ 10\%, and $f_0$ 
changes by no more than a factor $\approx 3$. 
We then proceed as follow. We calculate from the four fits given in
Table \ref{tab2} the mean values $\langle{f}_0\rangle$ and $\langle{\gamma}\rangle$.
We do not try to model the possible errors on ${f}_0$ and $\gamma$ but 
we just consider as their appropriate uncertainty range the bracketing
values found in our four models. We find: ${f_0}=3.72\,(+1.52,-1.30)
\times10^{-8}$ Hz and ${\gamma}=-1.08\,(+0.03,-0.04)$. 
In the following we will compute $\langle{h}_0\rangle$ and $\sigma_{h_0}$.

\subsection{Black hole mass function}

\begin{figure}
\centerline{\psfig{file=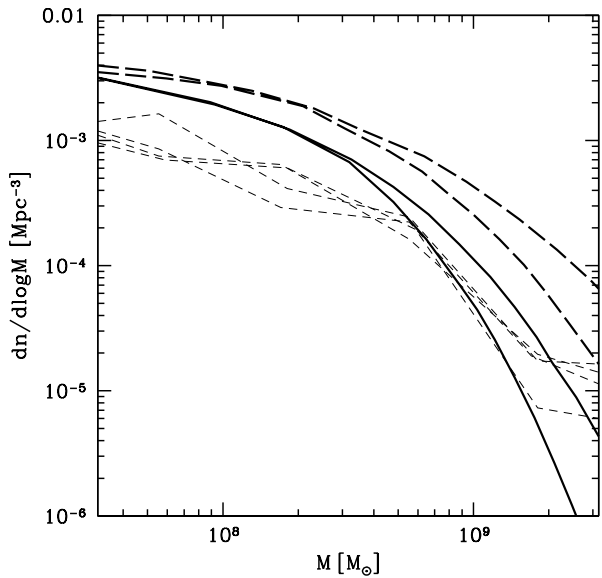,width=84.0mm}}
\caption{The local black hole mass function computed under different assumptions.
The thick lines bracket the uncertainties in the black hole mass function computed by Tundo et al. (2007) either on the basis of the $M_{\rm BH}$-$\sigma$ relation (solid lines), or assuming the $M_{\rm BH}-M_{\rm bulge}$ relation (dashed lines). The thin dashed lines are the outputs of the four merger tree models employed in this work.}
\label{figBHMF}
\end{figure}

Our poor knowledge of the shape and the normalisation of the black hole mass function (BHMF) at $z=0$ 
is the major uncertainty factor in the determination of ${\cal M}$.
The hierarchical models employed here are bound to reproduce 
at any redshift the $M_{\rm BH}-\sigma$ relation observed in the local universe (Tremaine et al. 2002).
The MBH population predicted at $z=0$ is then found to be consistent with the 
BHMF derived observationally by inferring MBH masses from measurements of
the bulge stellar velocity dispersion (Sheth et al. 2003).
The key point is that we can infer the BHMF either relying on the 
$M_{\rm BH}-\sigma$ relation, or on the $M_{\rm BH}-M_{\rm bulge}$ relation (e.g. Haring \& Rix 2004),
and the two approaches give inconsistent results especially 
for heavier MBHs (e.g Lauer et al. 2007). This is particularly relevant here
because the more massive MBHBs are the main contributors to the GW background
at very low frequencies (see Fig.~\ref{fig1} and Fig.~\ref{fig4}).
To quantify the uncertainties related to the BHMF we refer to the
analysis 
performed by Tundo et al. (2007). In Fig. \ref{figBHMF}
the local BHMFs (with the relative uncertainties) derived from the two
relations are shown (see Tundo et al. 2007 for details).
As it is clear from Fig. \ref{fig4}, almost all the signal comes from 
binaries with ${\cal M}>10^7$. The chirp mass is related to the primary (i.e. heavier)
black hole mass $M_1$ and the binary mass ratio $q=M_2/M_1\le 1$
by ${\cal M}=M_1\,q^{3/5}/(1+q)^{1/5}$. As a consequence, for a given $q$-distribution, the mean chirp mass
of the binaries is proportional to the mean mass of the MBH population.
Our analysis here concentrates on MBHs with masses 
$>10^{7.5}\msun$, that dominate
by far the GW stochastic background, see Fig.~\ref{fig1} and Fig.~\ref{fig4} . 
For each mass function we can calculate the average MBH mass of the population as
\begin{equation}
{M_{\rm av}}=\frac{\int_{7.5}^\infty \frac{dn}{d{\rm log}M_{\rm BH}}M_{\rm BH}\,d{\rm log}M_{\rm BH}}{\int_{7.5}^\infty \frac{dn}{d{\rm log}M_{\rm BH}}\,d{\rm log}M_{\rm BH}}\,.
\label{meanmass}
\end{equation}
Since in all our models the $q$-distributions have a similar shape, peaked
at $q \approx 0.1-0.2$, we can assume that, approximately, 
${\cal M}\propto M_{\rm av}$, so that the error on $M_{\rm av}$ directly 
quantifies the error on ${\cal M}$. We consider the four curves from Tundo 
et al. 2007 as they were the outcome of four different MBH formation models; 
we calculate $M_{\rm av}$ for each model, and then we evaluate the mean 
$M_{\rm av}$ and the relative variance.  We find
$\Delta {M_{\rm av}}/\langle{M}_{\rm av}\rangle\sim 0.31$.
Note that the uncertainty on ${M_{\rm av}}$ is dominated by the discrepancy
between the BHMF evaluated via the two different methods. The four merger tree
realisations that we have employed in this paper give results broadly consistent with the 
 BHMF determination using the $M_{\rm BH}-\sigma$ relation. The MBH mass density is in the 
range $2-3\times 10^5\msun$ Mpc$^{-3}$, consistent with the values 
$2.4-3\times 10^5\msun$ Mpc$^{-3}$ obtained integrating the BHMFs given 
by Tundo et al. 2007. The associated uncertainties in the BHMF at
$z=0$ are much smaller than those due to the dichotomy 
$M_{\rm BH}-\sigma/M_{\rm BH}-M_{\rm bulge}$, and we will neglect them here.

We note 
that the poor knowledge of the BHMF causes also an
uncertainty in the number density $n_{\rm BH}$ of MBHs at $z=0$;
focusing on the mass range relevant for this discussion, we define 
\begin{equation}
n_{\rm BH}={\int_{7.5}^\infty \frac{dn}{d{\rm log}M_{\rm BH}}\,d{\rm log}M_{\rm BH}}\,.
\label{meanmass}
\end{equation}
This fact is likely to produce an uncertainty in $N$. However we will see
in Section 5.2 that we can directly evaluate the uncertainty in
$N$ from the halo merger history, and we will also find that the associated 
uncertainties are much larger than those produced by the uncertainties in $n_{\rm BH}$;
the latter will therefore be ignored.

So far we have concentrated on the implications of the BHMF uncertainties
at $z=0$. However, Fig. \ref{fig3} shows clearly that all the MBHs coalescing
at $z<2$ contribute significantly  to the GW stochastic background. 
As our models need to reproduce the local $M_{\rm BH}-\sigma$ at any redshift, 
the possibility of a redshift evolution of this relation 
is a further factor of uncertainty in the background level.
Whether the $M_{\rm BH}-\sigma$ and $M_{\rm BH}-M_{\rm bulge}$ relations evolve with redshift is still a  matter of debate. 
The redshift evolution of the $M_{\rm BH}-\sigma$ is quite controversial.
Shields et al (2003) found that quasars between $z=0$ and 3.5 are 
consistent with the local $M_{\rm BH}-\sigma$ relation. Robertson
et al. (2006) investigated the galaxy merger driven $M_{\rm BH}-\sigma$ evolution with 
redshift by means of detailed hydrodynamical $N$-body simulations, finding almost no
evolution for $0<z<2$. On the contrary, Treu et al. (2004) found an offset of $\Delta {\rm log} 
\sigma =-0.16$ in quasars observed at $z=0.37$ with
respect to the local relation (Tremaine et al. 2002). If confirmed,
this would imply, for a given $\sigma$, an hosted MBH 4.4 times more massive at 
$z=0.37$ than in the local universe.
We can estimate the impact of the results by Treu et al. (2004) on the strength
of the GW stochastic background. Woo et al. (2006) provide the following
best fit for the offset with respect to the local relation:
\begin{equation}
\Delta {\rm log} M_{\rm BH}=1.66z+0.04 \,.
\label{msigmaoffset}
\end{equation}
If we correct the masses of the coalescing binaries according to
equation (\ref{msigmaoffset}) in the range $0<z<2$ and we decrease
accordingly the number of coalescing binaries in order to satisfy
the constraints on the MBH mass function given by the Soltan (1982) argument,  
$h_0$ would increase from $\simeq 10^{-15}$ to $\simeq10^{-14}$, 
that is on the border of the current PTA limits
(see Jenet et al. 2006 and the discussion in Section~\ref{s:obs}). This is because the characteristic amplitude
of the background is proportional to ${\cal M}^{5/3}N^{1/2}$, hence 
if we increase the mass and lower the number of sources by the same factor, 
the amplitude of the background increases. This is just an example of how PTAs could
provide information on the evolution of the MBH population
at low redshift.

There is growing evidence 
of a pronounced redshift evolution of the $M_{\rm BH}-M_{\rm bulge}$
relation. The best fit is given by McLure et al 2006:

\begin{equation}
\frac{M_{\rm BH}}{M_{\rm bulge}}=10^{-3.09}(1+z)^{2.07}\,.
\label{mBHmbulge}
\end{equation}
Given $M_{\rm bulge}$, this implies that ,e.g., the host MBH at $z=2$ should be eight time 
more massive than the host MBH at $z=0$. However at this stage
we cannot rigorously  quantify the implications of such possible 
evolution on the GW background, since
our models are based on the $M_{\rm BH}-\sigma$.
Robertson et al. (2006) found a significant redshift dependence in the
$M_{\rm bulge}-\sigma$ relation.
For a given $\sigma$ the corresponding $M_{\rm bulge}$ is four times
smaller at $z=2$ than now. This could be a clue that a mild redshift 
evolution in the $M_{\rm BH}-\sigma$ relation could be consistent 
with the evolution shown by equation (\ref{mBHmbulge}) in the
$M_{\rm BH}-M_{\rm bulge}$ relation. 

The redshift dependence of the $M_{\rm BH}-\sigma$ and $M_{\rm BH}-M_{\rm bulge}$
relations is currently debated and there is much freedom at 
present for theoretical speculations not constrained by observations. We have broadly quantified the possible implications on the strength of the background, depending on the actual scenario and we will not dwell on this any further. We just point out that PTAs -- even upper-limits on and not just detection of the stochastic background generated by populations of MBHBs -- will be able to provide direct new constraints on MBH populations and their assembly history that are likely to be difficult to infer using a number of indirect observations.

\subsection{Galaxy merger rate}

The major factor of uncertainty about the total number of sources
is related to the determination of the merger rate of massive 
(i.e. with total stellar mass $M_\star>10^{10}\msun$) 
galaxies at low redshift (say, $z\simlt 2$): 
once a relation between the MBH and the spheroid is assumed
(either the $M_{\rm BH}-M_{\rm bulge}$ or the $M_{\rm BH}-\sigma$ relation, as we have discussed in the last section),
the number of MBH pairs is proportional to the spheroid merger rate. 
EPS--based models tend to overpredict the bright end of the 
quasar luminosity function 
(LF) at $z<1$ (e.g. Marulli et al. 2006); however, although the
quasar activity is related to merger events, this discrepancy in
the predicted and observed LF 
does not automatically imply an overestimate of the 
galaxy merger rate. In fact little is known about the details of the
quasar powering mechanism, the accretion efficiency and 
quasar duty cycle, and the LF overprediction 
could be ascribed
to these factors, without necessarily implying an overestimate of the
galaxy merger rate.

A more robust test is to compare the low redshift
merger rates derived with EPS trees, with the merger rates
inferred by observations of the fraction of close galaxy pairs.
Our analysis is based of the most recent and complete study to date 
(Bell et al. 2006).
Starting from the COMBO survey data (Borch et al. 2006), Bell et al. 
give the mean fraction of pairs $\phi$ -- defined as galaxies with 3-D separation less than 30 kpc -- 
and the comoving number density of galaxies $n_\mathrm{G}$ in the redshift range $0.4<z<0.8$ assuming two different
galaxy total stellar mass cuts $M_{*,{\rm th}}$: (i) $M_{*,1,2}>2.5\times 10^{10}\msun$,  and (ii) $M_{*,1}>3\times 10^{10}\msun$ and  
$M_{*,2}>M_{*,1}/3$. Here $M_{*,1}$ and $M_{*,2}$ are the total masses of the primary and
secondary galaxy in the pair, respectively.

The merger rate of galaxies, assuming a typical merger timescale 
${\cal T}_{\rm M}$, is given by 
\begin{equation}
\dot{N}_{\rm M}=\frac{\phi}{2{\cal T}_{\rm M}} \int n_\mathrm{G}\, dV_c\,,
\label{obs_rate}
\end{equation}
where the integral is calculated between the survey redshift limits 
($0.4<z<0.8$) and the factor $1/2$ takes into account that two objects 
participate in a single merger event. Applying this calculation
to the two samples considered by Bell et al. 2006 and relying on their estimate of  
${\cal T}_{\rm M}\sim 4 \times 10^8$ yr (see also Barnes \& Hernquist 1992, Patton et al. 2002,
Lin et al. 2004, Naab Khockfar \& Burket 2006), we find $\dot{N}_{\rm M}\approx 5\times10^{-3}$ yr$^{-1}$
for case (i),  and $\dot{N}_{\rm M}\sim 2\times10^{-3}$ yr$^{-1}$ for case (ii).
Bell at al. estimate an error of $\sim 2$ in their results.
If we assume for massive galaxies a MBH occupation fraction of $1$
(e.g Richstone et al. 1998), the MBHB coalescence rate inferred 
by observations is the same as the galaxy merger rate.
To compare these values to the MBHB coalescence rate obtained from  
EPS based merger trees, we need to convert the galaxy stellar mass
threshold to a MBH mass threshold $M_{\rm BH,th}$. As we are dealing with
massive galaxies, to a first approximation we can use the $M_{\rm BH}-M_{\rm bulge}$
relation (Haring \& Rix 2004)
\begin{equation}
M_{\rm BH}=0.0014 M_{\rm bulge}\,,
\label{mBHmB}
\end{equation}
that has been shown to approximately hold at least up to $z=1$. McLure et al. 2006
found $M_{\rm BH}/M_{\rm bulge}\propto (1+z)^2$, but the normalisation of their best fit
to this relation at $z=0$ is lower than 0.0014; the relation given by equation
(\ref{mBHmB}) is within the error bars given by McLure et al. 2006 for $0.4<z<0.8$. 
Note that by adopting the relation~(\ref{mBHmB})  
we are implicitly assuming that all the observed galaxies are bulge-dominated. 
We can then consider only MBH more massive than $M_{\rm BH,th}$ according to equation~\ref{mBHmB}; if we integrate in the redshift range $0.4<z<0.8$ the MBHB merger rate obtained by EPS models, we obtain the values listed in the 
left hand columns of Table \ref{tab1}. The mean rates are a factor
3 lower than those inferred by applying equation (\ref{obs_rate})
to observations, and the dispersion around the mean is larger
for the case (ii). However, the $M_{\rm BH}-M_{\rm bulge}$ conversion
is too simplistic. At the considered mass cuts, the luminosity
function is dominated by disk type galaxies (Borch et al. 2006, 
Ilbert et al. 2006). 
Since typically only $\approx10\%$ of the stellar content of disk galaxies is contained in the bulge, 
this means that 
galaxies corresponding to the survey stellar mass threshold actually contain
a MBH with $M_{{\rm BH},{\rm th}}=0.00014 M_{*,{\rm th}}$, if we require them to fit
the relation (\ref{mBHmB}). For example a stellar mass cut of $M_{*,1}=3\times 10^{10}\msun$
corresponds to a host MBH of mass $\approx 4\times10^6\msun$ and not $\approx 4\times10^7\msun$. 
We then evaluate the MBH coalescence rate given by EPS trees
lowering the MBH threshold by an order of magnitude. By
doing so, we obtain the merger rates listed in the right hand columns
of Table \ref{tab1}; the mean rates are consistent -- they differ by
a factor $\lsim 2$ -- with those obtained from observational estimates. 
The EPS halo mass function is quite sensitive at high halo masses; 
we checked that changes in the mass function within 
reasonable limits affect the merger rate by $\lsim 40\%$. 
Since Bell at al. 2006 quote a factor of two uncertainty in their rate estimate we can safely conclude that their MBHB coalescence rates are consistent
with our EPS based theoretical models. 

Though Bell et al. 2006 give directly the merger rate of galaxies
 above a chosen mass threshold, there are several 
other works that evaluate the merger 
rate of galaxies in a given magnitude range. The comparison with 
the EPS merger tree estimates used in this paper is 
more difficult in this case, since we have to convert the magnitude 
limit into a mass limit which introduces a whole new range of
uncertainties. However, it is still valuable to compare the merger rates
obtained in this way with those derived with EPS merger trees as a sanity check. 
In order to do so, we combine the relations 
\be
{\rm log} L_H=0.8-0.5 M_B,
\label{LM}
\ee 
and
\be
{\rm log} M_*=\frac{{\rm log} L_H -0.12}{0.99},
\label{ML}
\ee
where $L_H$ is the $H$-band luminosity and $M_B$ in the $B$-band absolute 
magnitude (Zibetti et al. 2002).
Lin et al. (2004) derive an average merger 
rate of $4\times 10^{-4}$ $h_{100}^3$ Mpc$^{-3}$ Gyr$^{-1}$ 
at $0.5<z<1.2$ for galaxies
with $-21<M_B<-19$ (where $h_{100}$ is $H_0/100$ km s$^{-1}$ Mpc$^{-1}$ and we adopt $h_{100}=0.7$). 
If we multiply this number by the comoving volume enclosed
in the redshift range $0.5<z<1.2$ we obtain a merger rate of
$2.5\times10^{-3}$ yr$^{-1}$. Converting the magnitude limits in stellar 
mass thresholds using equations (\ref{LM}) and (\ref{ML}) 
and estimating the MBH mass using equation (\ref{mBHmB}), we obtain, from our
merger trees, rates of the order of $\sim 5\times10^{-4}$ yr$^{-1}$. 
If we use instead $M_{{\rm BH},{\rm th}}=0.00014 M_{*,{\rm th}}$, the rates
reach values $\sim10^{-2}$ yr$^{-1}$. Also in this case the observational
rate is consistent (at least within a factor of $\approx 5$) 
with the merger tree estimations.
Repeating the same analysis using the rate found by De Propris et al. (2007)
for galaxies with $-21<M_B<-18$ at $0.01<z<0.123$, we obtain once again
consistent results. Merger tree rates are found to be slightly higher (by a factor
$\lsim 2$) than the observed rates at very low redshift.
\begin{table}
\begin{center}
\begin{tabular}{c|cc|cc}
\hline
\multicolumn{1}{c|}{} & \multicolumn{4}{c|}{$\dot{N}_{\rm M}$}\\
\hline
\multicolumn{1}{c|}{} & \multicolumn{2}{c|}{$M_{\rm BH,th}=0.0014M_{*,{\rm th}}$} & \multicolumn{2}{c|}{$M_{\rm BH,th}=0.00014M_{*,{\rm th}}$}\\
\hline
$MODEL$ & (i) & (ii) & (i) & (ii)\\
\hline
   VHM&         $1.4\times10^{-3}$& $1.1\times10^{-4}$& $4.9\times10^{-3}$& $5.3\times10^{-4}$\\
   VHMhopk&     $2.4\times10^{-3}$& $2.0\times10^{-3}$& $9.4\times10^{-3}$& $5.1\times10^{-3}$\\
   KBD&         $2.1\times10^{-3}$& $2.2\times10^{-4}$& $9.6\times10^{-3}$& $1.1\times10^{-3}$\\
   BVRhf&       $1.8\times10^{-3}$& $3.4\times10^{-4}$& $1.2\times10^{-2}$& $1.5\times10^{-3}$\\
\hline
\end{tabular}
\end{center}
\caption{Merger rates inferred from our EPS merger trees in the redshift
range $0.4<z<0.8$. Columns (i) assume a stellar mass cut
$M_{*,1,2}>2.5\times 10^{10}\msun$, while columns (ii) consider
$M_{*,1}>3\times 10^{10}\msun, M_{*,2}>1/3M_{*,1}$. Results are shown
by assuming that all the observed galaxies are elliptical 
($M_{\rm BH,th}=0.0014M_{*,{\rm th}}$), and by taking into account 
that the galaxy sample at the threshold mass is dominated by spirals
($M_{\rm BH,th}=0.00014M_{*,{\rm th}}$).}
\label{tab1}
\end{table}

Since the differences between model predictions and observations are of the same order of the spread of the merger rate values that we find in our four models, we directly estimate the mean square range of MBHB merger rate $\sigma_{\dot{N}_{\rm M}}$ using the rate shown in Table \ref{tab1}; we find $\sigma_{\dot{N}_{\rm M}}/ \langle{\dot{N}}_{\rm M}\rangle\simeq 0.76$.
Since ${\cal M}\propto M_{\rm av}$ and $N\propto \dot{N}_{\rm M}$,
we can write
 \begin{equation}
{\left(\frac{\sigma_{h_0}}{\langle h_0 \rangle}\right)^2=\left(\frac{1}{2}\frac{\sigma_{M_{\rm av}}}{\langle{M}_{\rm av}\rangle}\right)^2+\left(\frac{5}{3}\frac{\sigma_{\dot{N}_{\rm M}}}{{\langle{\dot{N}}}_{\rm M}\rangle}\right)^2.}
\label{error2}
\end{equation}
Substituting $\sigma_{\dot{N}_{\rm M}}/ \langle{\dot{N}}_{\rm M}\rangle= 0.76$ and 
$\sigma_{M_{\rm av}}/\langle{M_{\rm av}}\rangle= 0.31$ into equation~(\ref{error2}), we find
\begin{equation}
\frac{\sigma_{h_0}}{{\langle h_0 \rangle}} = 0.65.
\label{errorA}
\end{equation}
We need now to determine $\langle{h_0}\rangle$ and we proceed as follows. The merger trees are bound to reproduce the $M_{\rm BH}-\sigma$ relation, so they provide a 
value of $\langle{h_0}\rangle$ that is not representative of the whole range of possibilities 
for the BHMF shown in Fig.~\ref{figBHMF}. To provide a more representative estimate of
 $\langle{h_0}\rangle$, we calculate $\langle{h_0}_{(M_{\rm BH} - \sigma)}\rangle$ from our merger tree models, and then we correct it for the fact that the $M_{\rm av}$ inferred from the $M_{\rm BH}-\sigma$ relation ($M_{{\rm av},(M_{\rm BH} - \sigma)}$) 
is lower than $M_{\rm av}$ if also the $M_{\rm BH}-M_{\rm bulge}$ relation is taken into account:

\begin{equation}
\langle{h_0}\rangle=\langle{h_0}_{(M_{\rm BH} - \sigma)}\rangle\left(\frac{M_{\rm av}}{M_{{\rm av},{(M_{\rm BH} - \sigma)}}}\right)^{5/3}\,.
\label{Amean}
\end{equation}
(Here we used the fact that, for a given mass ratio distribution, the mean ${\cal M}$ is proportional to $M_{\rm av}$.)
Since $M_{\rm av}/M_{{\rm av},{(M_{\rm BH} - \sigma)}}\simeq 1.4$ (from Tundo et al. 2007) and 
$\langle{h_0}_{{(M_{\rm BH} - \sigma)}}\rangle\simeq 1.1\times 10^{-15}$, we finally find
${h_0}=(1.93\pm1.25)\times10^{-15}.$

\begin{figure}
\centerline{\psfig{file=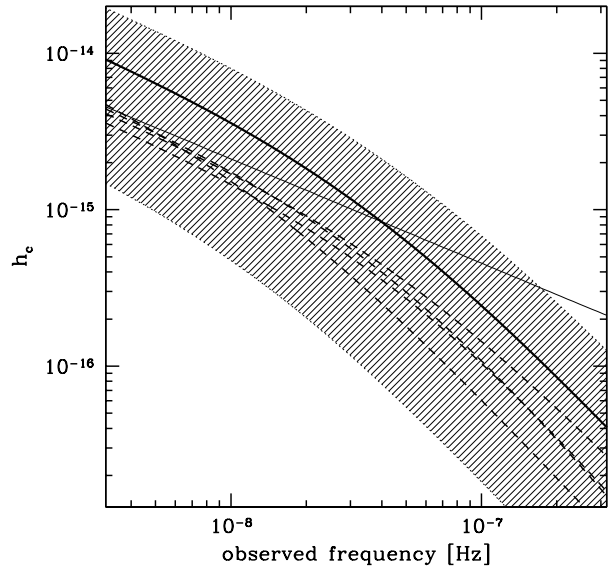,width=84.0mm}}
\caption{Summary of the uncertainties of the predicted characteristic amplitude of the GW stochastic background. 
The thick solid line shows the mean level of the signal 
according to the 
analysis 
discussed in the text, while the thin dashed lines represent the predictions of the four MBHB assembly scenarios considered in this paper.
The shaded area delimits the $1\sigma$ region around the mean value. For comparison,
the na\"{\i}ve power law $f^{-2/3}$ is plotted as a thin solid line for the VHM model.}
\label{figunc}
\end{figure}

Summarising, the GW stochastic background is 
well approximated by the relation given 
by equation~(\ref{hfit}) that depends on three parameters
whose range of values is
\begin{eqnarray}
\label{fitparameters1}
{h_0} & = & (1.93\pm1.25)\times10^{-15}\,,
\label{fitparameters}
\\
{f_0} & = & 3.72^{+1.52}_{-1.30} \times 10^{-8}\,\mathrm{Hz}\,,
\label{fitparameters2}
\\
\gamma & = & -1.08^{+0.03}_{-0.04}\,;
\label{fitparameters3}
\end{eqnarray}
their specific value for the four models
considered in this paper are shown in Table~\ref{tab2}.  
The results of our analysis are shown in Fig. \ref{figunc}. 
As expected, the outcomes of our merger trees, based on the
$M_{\rm BH} - \sigma$ relation, lie in the lower region of the estimated 
error bar, below the best fit to the background.

\subsection{Other uncertainty factors}

The spread of values of the estimate of GW stochastic background
that we have discussed above are likely 
to be affected by other factors, in particular 
the MBHB coalescence efficiency, the gravitational recoil and the MBH accretion prescription adopted 
during mergers. We 
consider them now in some detail.
 
Our knowledge of the MBHB coalescence efficiency
is far from robust. The major contribution to the GW background signal comes
from massive binaries at low redshift, that are expected to be hosted
in early type, gas poor spheroidal galaxies. Stellar dynamical processes
are therefore likely to be the main driver in MBHB coalescences.
It is however known that 3-body interactions (that extract energy from the MBHBs through the so called slingshot mechanism) are not sufficient to lead
to a MBHB merger within an Hubble time in spherical systems 
(e.g. Sesana, Haardt \& Madau 2007b),
unless the binary is very eccentric. The
scenario is different in
axisymmetric and triaxial systems (more realistic if we consider
the end-product of a merger), and there are indications (Berczik et al. 2006) 
that in the latter case the stellar supply to the slingshot process
could be sufficient to drive binaries to coalescence within an Hubble time. If only a fraction $\epsilon_c$  of binaries coalesce, then 
the stochastic background characteristic amplitude $h_c$ drops by a factor $\sqrt{\epsilon_c}$ at all frequencies. 

A MBH formed via the merger of a binary system recoils due to the net
linear momentum imparted on the object through emission of GWs 
(Redmount \& Rees 1989). The recoil velocity has been
shown to possibly reach high values $\sim 10^3$ km s$^{-1}$ (e.g. Campanelli et al. 2007, Tichy \& Marronetti 2007) and 
may therefore lead to a depletion of MBHs in the core of some galaxies. These galaxies involved in mergers at later cosmic times would therefore not form a MBH binary, with a consequent decrease in the number of MBHBs and in the strength of the associated
GW stochastic signal. In order to address the impact of GW recoil on the
estimates of the level of the stochastic background, we have run a series of tests using
the VHMhopk and BVRhf models where we compare the signal strength in the case where the
recoil is neglected and the
case where the recoil is included using an {\em extreme} prescription following 
Campanelli et al. 2007.
We find negligible differences in the BVRhf case, whereas 
differences as large as a factor of two in the GW background amplitude occur in the VHMhopk scenario. 
However such a difference is between the two  
prescriptions of no recoil at all and of extreme recoil. 
We know that the GW recoil exists, and the differences in the computed
background between the non spinning MBH recoil recipe 
(i.e. our default models) and the extreme recipe for maximally 
spinning MBHs is indeed tiny.

The modeling of accretion onto a MBH could also influence the background level.
Although we have tested that the GW background is basically unaffected
by the details of the accretion onto a single MBH 
({\em e.g.} the dependence on $z$ and spin),  
an important issue is whether both MBHs typically accrete mass during
mergers. We have considered models in which both
MBHs are subject to mass accretion (in our original
models only the more massive BH in the pair undergoes such a stage), finding that the
predicted background amplitude could even double, due to the increase of
the MBHB chirp mass. 

All these factors add further uncertainties to the estimates reported in the 
previous section, but our knowledge of the MBH accretion and
the coalescence efficiency are too poor at present to allow us to provide
stringent quantitative constraints on the level of the GW background. In general,
the uncertainties due to the accretion prescription should change by at most
a factor $\approx 2$ the amplitude of the signal, and the coalescence 
efficiency could just reduce the strength of the background with respect to the
values reported here, since in all our models we have set $\epsilon_c = 1$.
    
\section{Comparison with previous work}

The  very low-frequency GW stochastic 
background from MBHBs has been computed in a handful of
papers over the last ten years. Previous works have only concentrated on a (in turn
different) {\em specific}
MBHB assembly scenario
, none of which matches exactly any of those considered in this paper; they
have modeled the background as a simple
power law as given by equation~(\ref{hcpar}), that we have shown does not correctly
describe the signal across the whole frequency band accessible to PTA observations. 
In this section we briefly summarise how previous estimates compare among each other
and with those presented in this paper.

Studies based on the EPS formalism 
(Wyithe \& Loeb 2003, Sesana et al. 2004, Enoki et al. 2004),
though relying on different prescriptions for the seed BH mass, the 
MBH accretion, and the evolution of the stellar spheroids, yield all
$h_\mathrm{1yr}\sim 10^{-15}$, see equation (\ref{hcpar}).
This is not surprising, since the very low frequency background
is strongly dominated by massive nearby sources that are
fairly constrained by current observations (see the discussion
in the previous section): there are well established 
relations between the BH mass, the spheroids mass 
(Magorrian et al. 1998) and the velocity dispersion (Ferrarese \& Merritt 2000, 
Gebhardt et al. 2000, Tremaine et al. 2002) 
that have been shown to hold at least up to $z\lsim 1$ (McLure \& Dunlop 2002,
McLure et al. 2006, Peng et al. 2006), and the
stellar and nuclear MBH mass density are today known within an uncertainty of a factor $\approx 2$ (e.g. Marconi et al. 2004, Borch et al. 2006). 
Since the goal of every assembly model is to reproduce these low 
redshift observational constraints, all the scenarios
based on the same halo merger rate function (extracted in this 
case using the EPS formalism, either with analytical or  
numerical techniques),  lead to comparable (i.e. within a factor $\approx 2$) 
low frequency GW stochastic background levels.
\begin{figure}
\centerline{\psfig{file=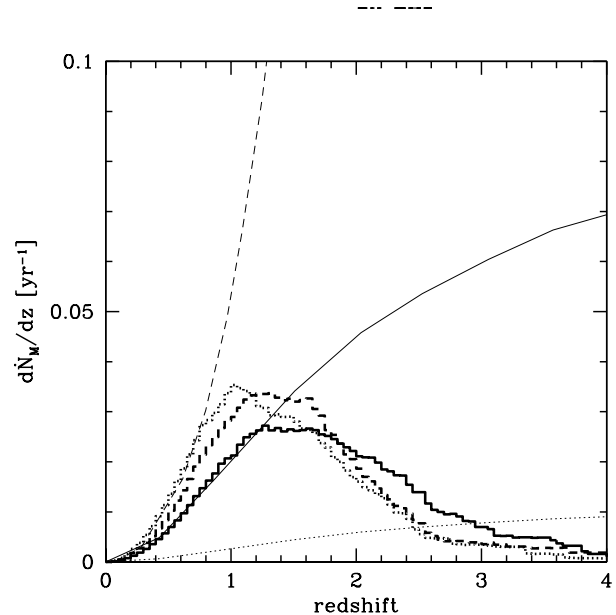,width=84.0mm}}
\caption{Spheroid merger rates and massive black hole binary coalescence rates (per unit redshift) as a function of redshift. The histograms represent the coalescence rates for binaries with ${\cal M}>10^{7.5}\msun$ predicted by the VHM (thick solid line), KBD (thick dotted line) and BVRhf (thick dashed line) models. The thin solid line shows the spheroid merger rate adopted by J\&B, while the thin dotted line is the coalescence rate for MBHBs with ${\cal M}>10^{7.5}\msun$ that we have derived from the same model considered by J\&B. The thin dashed line correspond to the galaxy merger rate adopted by R\&R.}
\label{mrate}
\end{figure}

Earlier work (R\&R, J\&B), that inferred the MBHB coalescence rate 
from available observations of the fractions of galaxies 
forming kinematically close pairs in the local universe, give somewhat
lower (by a factor $\sim 5$) normalisations of the spectrum.   
It is useful to recall here that these models  
focus specifically on the GW background, and do not try to match other
observational constraints in a self--consistent MBH evolution model.
J\&B used the spheroid merger rate found by Carlberg et al. 2000
for $0.1<z<1.1$, and then populate the spheroids with MBHs according 
to a log-normal MBH mass function given by 
\begin{equation}
\left<{\rm log}_{10}\frac{M_{\rm BH}}{\msun}\right>={\rm log}_{10}(1.2\times 
10^7)\pm 0.6\,.
\label{mdistJB}
\end{equation}
However, the merger rates of Carlberg et al. 2000 are evaluated for galaxies
with magnitude $M_B<-20.4$, corresponding to massive galaxies with 
stellar mass $M_*\gsim 2\times 10^{10}\msun$. In practice J\&B
use the merger rate of massive galaxies as the {\it total} galaxy merger rate.
In Fig. \ref{mrate}, the galaxy merger rates used by J\&B and R\&R are compared
with the MBHB coalescence rates obtained with our EPS based 
halo merger tree models placing a lower cut at ${\cal M}=10^{7.5}\msun$.
EPS rates for ${\cal M}>10^{7.5}\msun$ are similar to those used by both
J\&B and R\&R as {\it total} rates at $z<1$, reflecting the fact that those
'observational' rates refer to massive galaxies. Using the MBH mass
function distribution given by equation (\ref{mdistJB}) we find that
only $\approx 13\%$ of the J\&B mergers involve binaries with   
${\cal M}>10^{7.5}\msun$, 
so the discrepancy of a factor of $\approx 5$ in the
expected signal is reasonable. Similar considerations hold for the
R\&R model, which moreover employs very conservative assumptions on
the MBH number density evolution with redshift.  

\section{Observational consequences}
\label{s:obs}

\subsection{Pulsar Timing Arrays}

We have considered a wide range of models of formation and evolution of
MBHBs and estimated the GW stochastic background that one can expect
as a function of the model parameters. We have further shown that the 
stochastic signal does not follow a simple power-law $h_c(f) \propto f^{-2/3}$ 
for $f\simgt 10^{-8}$ Hz, as previously claimed. These results
are summarised in equation (\ref{hfit}), Table~\ref{tab2} and Fig. \ref{figdrop}. We can now explore, for a
much more realistic range of models than those considered in the past,
the observational consequences; in particular we can investigate whether current and/or future
PTAs would be able to detect such a signal, opening therefore
a new window for the study of MBHBs and structure formation, or at least setting
sufficiently stringent upper-limits to rule out selected scenarios. 

\begin{figure}
\centerline{\psfig{file=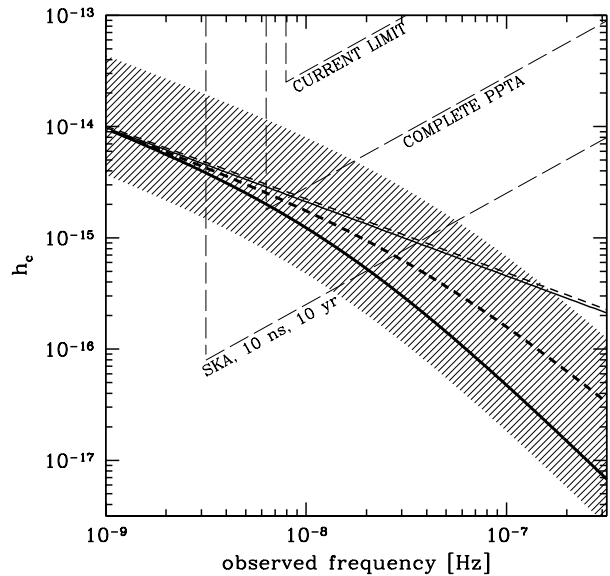,width=84.0mm}}
\caption{The sensitivity of Pulsar Timing Array in observations of a GW stochastic background. The long--dashed thin lines represent the sensitivities of current and future timing experiments, as labeled in the figure (see the text for more details). Three cases are considered: the current limit and the estimated sensitivity achievable by monitoring 20 pulsars for 5 years with $\delta t_\mathrm{rms} = 100$ ns -- complete Parks PTA (PPTA) -- are taken from Jenet et al. (2006). The sensitivity achievable with the Square Kilometer Array -- assuming the monitoring of 20 pulsars for 10 years at a precision level of $\delta t_\mathrm{rms} \sim 10$ ns -- is also shown. The solid lines depict the stochastic signal predicted by the na\"{\i}ve semi-analytical approach (thin), see equation~(\ref{hcpar}) and by our Monte-Carlo approach (thick), see equation~(\ref{hfit}) and Table~\ref{tab2}, assuming the VHM model for the massive black holes cosmic evolution; the short dashed lines refer to the KBD model. The shaded area marks the possible range of GW background level, according to the model uncertainties, as discussed in Section 5.}
\label{detection}
\end{figure}

As we have discussed in the Introduction, the technique to search for a stochastic background using PTAs consists in correlating the timing residuals from $N_P$ pulsars\footnote{We note that an upper-bound to any stochastic GW background can be placed by monitoring a {\it single} pulsar and assuming that the observed value of the timing residuals is entirely due to gravitational waves.}. The timing residual from a given pulsar is given by
\be
\delta t = \delta t_n + \delta t_h
\ee
where $\delta t_h \sim h /f$ and $\delta t_n$ are the GW signal and noise contribution, respectively. If one
considers for simplicity the case of only two pulsars, $N_P = 2$, then the minimum detectable signal (assuming that its spectrum is dominated by the low frequency contribution, as it is the case here) is characterised by:
\begin{equation}
h_{100}^2\Omega_\mathrm{gw}(f)\propto\frac{\delta t_\mathrm{rms}^2f^4}{\sqrt{T\Delta{f}}},
\label{omega}
\end{equation}
where $\delta t_\mathrm{rms} = \sqrt{\m \delta t^2\M}$ is the root-mean-square value of the timing residuals and
$\Delta f$ the bandwidth of the search. Note that this is the PTA equivalent of the result that one obtains by 
considering {\em direct} searches for GW stochastic backgrounds using the cross-correlation of the data from two GW laser interferometers (see Allen \& Romano, 1999 for a detailed discussion, and in particular Section III for material relevant to this section). If instead of two, one has many pulsars, then the optimal signal-to-noise ratio (SNR) is given by the combination of all the statistically independent correlations that can be formed:
\begin{equation}
{\rm SNR}_{\rm T}^2=\sum_{i<j}^{N_P} {\rm SNR}^2_{ij}=\frac{N_P(N_P-1)}{2}{\rm SNR}^2,
\label{snr}
\end{equation}
where ${\rm SNR}_{ij}$ is the SNR of the pair composed by the $i$-th and the $j$-th pulsars, and in the second equality we have assumed for simplicity ${\rm SNR}_{ij}={\rm SNR}$ $\forall i,j$. For $N_P$ sufficiently large, the signal-to-noise ratio scales as ${\rm SNR}^2_{\rm T}\propto N_P^2$ and therefore:
\begin{equation}
h_{100}^2\Omega_\mathrm{gw}(f)\propto\frac{\delta t_\mathrm{rms}^2f^4}{N_P\sqrt{T\Delta{f}}}.
\label{omega2}
\end{equation}
Using equations~(\ref{e:Omega}) and~(\ref{pinnei}) we finally obtain:
\begin{equation}
h_c(f)\propto\frac{\delta t_\mathrm{rms}f}{N_P^{1/2}(T\Delta{f})^{1/4}}.
\label{hcpulsar}
\end{equation}
The PTA sensitivity scales as $h_c(f)\propto f$ and reaches a minimum detectable frequency $f\sim 1/T$. This produces the characteristic wedge-like sensitivity curves as those shown in Fig. \ref{detection}. The sensitivity is proportional to the timing precision ($\delta t_\mathrm{rms}$) and improves as the square root of the number of observed pulsars; the improvement with the observational time {\it at a given frequency} scales as $h_c(f)\propto T^{-1/4}$, but note that a longer $T$ implies that lower frequencies can be reached, which provides an additional sensitivity gain. The curves shown in Fig. \ref{detection} are derived by normalising equation (\ref{hcpulsar}) to the sensitivity quoted by Jenet et al. (2006) for the complete Parkes PTA 
(labelled as "complete PPTA" in Fig. \ref{detection}), 
defined as the monitoring of 20 radio-pulsars for 5 years with $\delta t_\mathrm{rms} = 100$ ns: this
array would provide a peak sensitivity of $h_c \approx 2\times 10^{-15}$ at $f = 7\times 10^{-9}$ Hz.
Jenet et al. (2006) quote for the current Parkes PTA (labelled as "current limit" in Fig. \ref{detection}) an
optimal sensitivity equivalent to $h_c \approx 2\times 10^{-14}$ at $f = 8\times 10^{-9}$ Hz.  Fig. \ref{detection} clearly shows
that such a sensitivity is a factor of $\sim3$ above the most optimistic estimate of the background that 
we have considered. No meaningful astrophysical information about massive black hole binaries can therefore
be derived from the current upper-limit.
However, the complete Parkes PTA should provide a sensitivity improvement of about an order of magnitude in amplitude; several scenarios and/or parameter values within a given model could yield a detection at this level; if detection is not achieved, upper-limits could start ruling out, for example, scenarios in which the $M_{BH}-M_{bulge}$ relation holds for BHs of all masses at all redshifts.However, the sensitivity of the PTA decreases in the frequency range where the stochastic signal is expected to deviate from the $f^{-2/3}$ behaviour, and positive detection is therefore not guaranteed. The planned SKA will provide a major increase in PTA capabilities. Assuming that the SKA will be able to monitor 20 millisecond pulsars for ten years at a precision level of $\delta t_\mathrm{rms} \sim 10$ ns, it will allow us not only to detect the GW stochastic background, but also to study its spectrum in the frequency range $3\times10^{-9} \lsim\, f\, \lsim\,{\rm few}\times10^{-8}$, enabling the complete characterization of the GW signal. In turn, this would allow us to derive detailed information about the MBH formation and evolution history. Even considering a less optimistic timing precision of $\delta t_\mathrm{rms} \sim 50$, a comparable sensitivity can be obtained by monitoring $\approx 100$ millisecond pulsars.

\subsection{Complementarity of {\it LISA} and PTAs}

\begin{figure*}
\centerline{\psfig{file=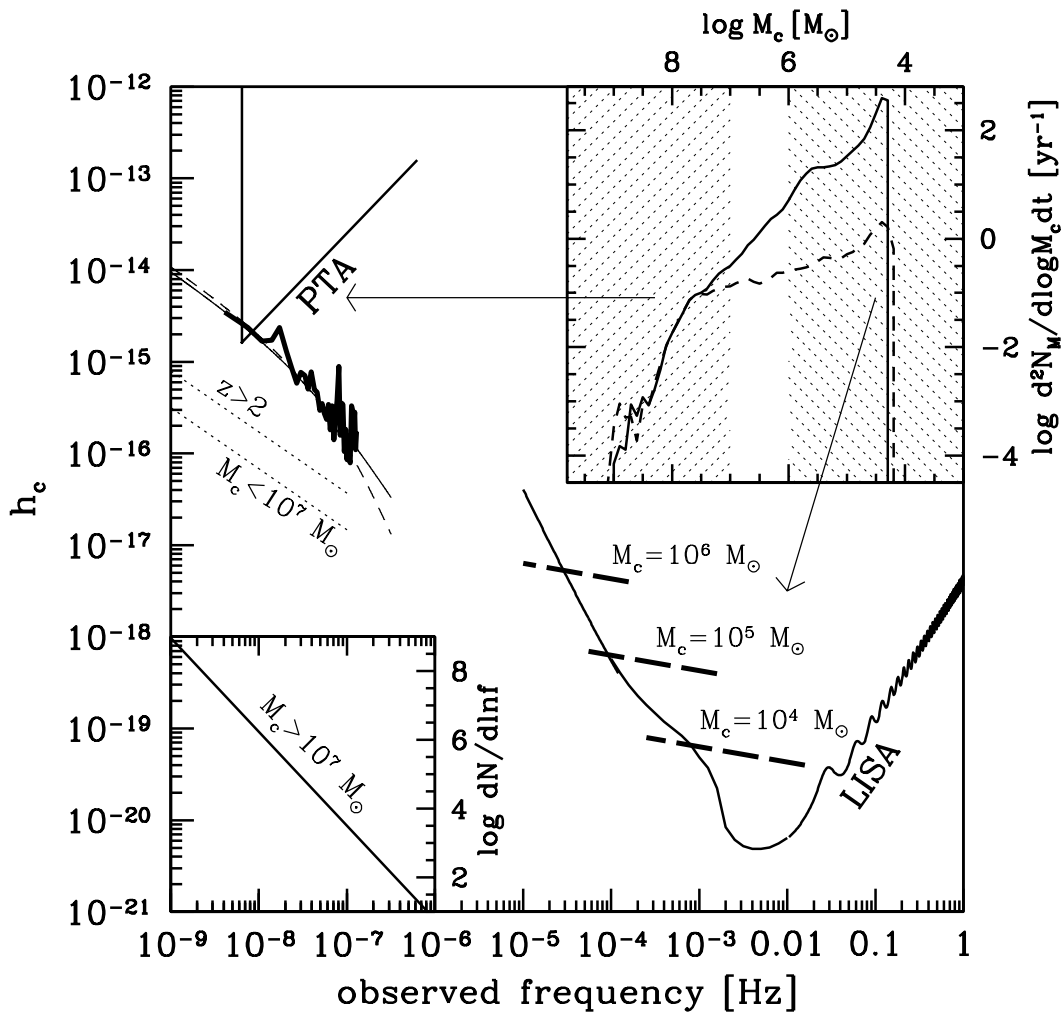,width=144.0mm}}
\caption{A representative summary of GW observations of the cosmic history of massive black hole binaries with Pulsar Timing Arrays (through the detection of a GW stochastic background) and {\it LISA} (via the direct detection the final stage of the coalescence of individual binary systems). In the main panel we show the characteristic amplitude of GW signals as a function of frequency, compared to the sensitivity of the complete Parkes Pulsar Timing Array and {\it LISA}. The thin lines in the main panel represent the GW stochastic background 
(according to equation~(\ref{hfit}) and Table \ref{tab2}) for a massive black hole formation model in which seeds are abundant (KBD model, solid line) and one in which seeds are rare (BVRhf model, dashed line). A specific Monte-Carlo realization  of the total signal at frequencies accessible to Pulsar Timing Arrays is the thick jagged line generated by nearby ($z < 2$) massive (${\cal M} > 10^7\,\msun$) sources: the thin dotted lines show in fact that the contribution by massive black hole binaries at $z>2$ or with ${\cal M}<10^7\msun$ is indeed negligible. In the {\it LISA} frequency band one will observe the coalescence of individual sources, and most of the events involve binaries with $10^4\msun<{\cal M}<10^6\msun$ at $z>3$ (the thick long-dashed lines show the trace of the final stage of the in-spiral for selected masses). The upper right panel shows the massive black hole binary coalescence rate $d^2N_M/d\log {\cal M}dt$ as a function of ${\cal M}$ for the two models KDB and BVRhf, which is the same for ${\cal M} \simgt 10^8\,\msun$ but differs as much as a factor $\sim 100$ for ${\cal M} \sim 10^4\,\msun$.  The lower left panel shows the number of sources with ${\cal M}>10^7\msun$ contributing to the GW background. The slope of the curve 
is such that the signal is stochastic at very-low frequencies, and that the {\it LISA} detection rate of binaries with ${\cal M}>10^7\msun$ is less than 1 yr$^{-1}$.}
\label{figskatch}
\end{figure*}

In Section 3.2, see Fig. \ref{fig1},  we have shown that very different models of MBH assembly 
(VHM, KBD, BVR, VHMhopk) predict a GW stochastic background whose 
characteristic amplitude varies by less than  factor of 2 at $f \sim 10^{-8}$ Hz, 
despite the radically different evolutionary paths of MBHBs and the number and mass of the initial seeds.
These very same models can also be used to estimate the total number of coalescences of MBHBs detectable by {\it LISA} (Bender et al. 1998) in the frequency range 0.1 mHz - 0.1 Hz, therefore  $\sim 5$ orders of magnitude higher than the frequency window probed by PTAs. In the
case of {\it LISA}, the numbers of coalescences of MBHBs differ by more than a factor 100. 
As a specific example, the BVRhf and KBD models predict $\sim 10$ and 
$\sim 10^3$ {\it LISA} events, respectively (Sesana, Volonteri \& Haardt 2007a), with MBHB coalescence 
taking place at frequency $\sim 1$ mHz. This is in
stark contrast with predictions for a GW stochastic background, whose strength shows
just a  $\approx 10\%$ difference in the nHz range. As we pointed out in Section 6, these results reflect the fact that
all our models 
reproduce by construction the local $M_{\rm BH}-\sigma$ relation, 
and being the main contributors to the very low frequency signal 
massive (${\cal M}\simgt 10^8\msun$), nearby 
($z \simlt 2$) binary systems, their population is fairly well constrained by this assumption. 
On the other hand, very different MBH formation scenarios are compatible with 
the observed $M_{\rm BH}-\sigma$; being the majority of the {\it LISA} sources 
lighter binaries at high $z$ (up to $z\gsim 10$), current observations do not place 
stringent constraints on their abundance, and there is much more room for speculation. 

{\it LISA} and PTA observations provide therefore fully complementary information
on MBHs and their formation history. As we have shown in Section 5, the level of the stochastic background
in the PTA frequency range is sensitive (a factor of $\approx$5) to the uncertainties in the MBH mass function at low redshift, 
to the halo merger rate variance, and to the accretion prescription, 
but is fairly insensitive to the nature and abundance of the MBH seeds.
In Fig. \ref{figskatch} we provide a summary sketch of such complementary.  
We consider the KBD and the BVRhf assembly models 
which produce GW backgrounds at very low frequencies 
whose characteristic amplitudes differ by less than $10\%$ (main panel). The models are 
characterised by mergers of low-mass MBHs at high redshift; the merger rates (shown in the 
upper-right panel as 
a function of chirp mass) differ by two orders of magnitude. The lower left box shows that while the low frequency background is composed by 
$\sim 10^3 - 10^6$ sources with ${\cal M}>10^7\,\msun$, they are basically absent in the {\it LISA} sensitivity window, with a detection rate $\lsim1$ yr$^{-1}$.

It is clear that PTA and {\it LISA} observations will place very different constraints on
the assembly history of MBHBs and the nature of the 
MBH seeds. It also follows that the detection of a GW background with PTAs will 
bound very weakly the expected number of MBHBs observable with {\it LISA}. 
It is equally straightforward to verify that the reverse is also true. The vast majority of the 
{\it LISA} detections is associated with the first round of coalescence events after the MBH seed formation
(Sesana 2007). The subsequent merger and accretion history of black holes has 
a minor impact, by a factor $\simlt 2$, on the number of events, once the mass and the abundance
of the seeds is defined (Sesana et al. 2007). It is not difficult to envisage two models with the
same seed MBH population but with different assumptions regarding, e.g., the accretion or the relation 
between the MBH and the host galaxy; such models
would predict similar {\it LISA} number counts but a different GW background level in the PTA frequency range.   
As an example we find that allowing accretion on both MBHs during each merger, instead of the standard
accretion onto the more massive one, could affect the very low frequency GW level by a factor of two, leaving 
essentially unchanged the {\it LISA} number counts.

\section{Summary}

We have carried out a systematic study of the GW stochastic background generated
by populations of MBHBs in the frequency range accessible to PTAs. We have considered a
number of models that qualitatively encompass the whole range of assembly scenarios proposed so far and quantitatively covers a broad spectrum of predictions for MBH formation. Our analysis shows
that, regardless of the actual model, the stochastic background is not described by a simple power law across the whole frequency spectrum (as considered so far) but by the super-position of two power-laws with a knee frequency around $10^{-8}$ Hz. Above such frequency the amplitude of the background decreases more rapidly than $f^{-2/3}$. This is a direct consequence of the discrete nature of sources. A phenomenological parametrisation of the amplitude of the GW stochastic background is given by equation~(\ref{hfit}). We have also carefully considered the present uncertainties that surround the actual values of the key parameters that set the level of the stochastic background and have provided an upper and lower limit (based on the current theoretical understanding and observational constraints) to the strength of such signal: this provides useful benchmarks for the planning of future PTA observations. 

As a byproduct of our analysis, we have compared 
the observed massive (luminous) galaxy merger rates at $z<1$ derived by a number of authors 
with the merger rates inferred using the EPS approach adopted in this paper. We find 
good agreement, with merger rates consistent with constraints provided by observations
when we adopt standard prescriptions for the conversion of the galaxy luminosity
into a MBH mass. 

We have shown that the estimated GW background level is close -- within a factor of $\approx 3$ for the most optimistic predictions -- to the upper-limit provided by current PTAs.
The complete Parkes PTA should be able to detect the background 
predicted by a number of assembly models and even upper-limits would already be able
to rule out selected scenarios, constraining the MBH formation history. 
The planned SKA observatory will detect the background
in a broad frequency band, including the steeper part of the spectrum.
The timing of 20 pulsars for 10 years with a $\delta t_\mathrm{rms} \simeq 10$ ns 
should allow us to study the frequency dependency of the GW background in the frequency range 
$3\times 10^{-9}-3\times 10^{-8}$ Hz, providing unique information about 
the MBHB demographics at low redshift ($z<2$).  However the data analysis approach 
-- as well as the number and stability of pulsars available for such studies -- 
will need to be considered in detail 
to assess precisely the sensitivity level achievable by any given
survey. This is beyond the scope of this paper and we have not tried to provide a detailed
discussion of these complex issues.

In the frequency range $\sim 10^{-8}\,\mathrm{Hz} - 10^{-7}\,\mathrm{Hz}$  the overall GW signal also shows prominent peaks generated by individual sources that are particularly bright and that may be resolved with PTA observations. We have not considered these signals here -- our work has been entirely focused on the GW stochastic contribution -- but they are
potentially of great interest; this is an area that deserves a systematic and quantitative study and we plan to return to this issue in the future.

The population of massive black hole binary systems is a primary target for both Pulsar Timing Arrays and the Laser Interferometer Space Antenna. We have shown that PTA observations of a GW stochastic background generated from such a population are ``orthogonal'' and complementary to {\it LISA} observations of individual events. PTAs and {\it LISA} will therefore {\em jointly} provide unique information about MBH assembly throughout cosmic history. 

\section*{Acknowledgments}

We would like to thank A. Lommen for useful discussions concerning searches for gravitational waves with Pulsar Timing Arrays and M. Volonteri for providing the Monte-Carlo realisations of the halo and MBH merger hierarchy. This work has been supported by the UK Science 
and Technology Facilities Council. While at Northwestern University, AV was supported by the David and Lucile Packard Foundation and by NASA grant NNG06GH87G.

{}

\end{document}